\documentclass[letter,longauth]{aa}

\usepackage{graphicx}
\usepackage{txfonts}

\usepackage{amsmath,bm}
\usepackage{amssymb}

\usepackage{lipsum}

\usepackage{natbib}
\usepackage{hyperref}

\newcommand{\RA}{\mathrm{RA}}
\newcommand{\DEC}{\mathrm{DEC}}
\renewcommand{\vec}[1]{\bm{#1}}

\begin{document}

   \title{The mass of $\beta$ Pictoris c from $\beta$ Pictoris b orbital motion
  }
\author{ S.~Lacour\inst{\ref{lesia},\ref{esog}}
 \and J.~J.~Wang\inst{\ref{caltech}} \thanks{51 Pegasi b Fellow.}
 \and L.~Rodet\inst{\ref{cornell}}
 \and M.~Nowak\inst{\ref{cam}}
 \and J.~Shangguan\inst{\ref{mpe}}
 \and H.~Beust\inst{\ref{ipag}}
 \and A.-M.~Lagrange\inst{\ref{ipag},\ref{lesia}}
 \and R.~Abuter\inst{\ref{esog}}
 \and A.~Amorim\inst{\ref{lisboa},\ref{centra}}
 \and R.~Asensio-Torres\inst{\ref{mpia}}
 \and M.~Benisty\inst{\ref{ipag}}
 \and J.-P.~Berger\inst{\ref{ipag}}
 \and S.~Blunt\inst{\ref{caltech}}
 \and A.~Boccaletti\inst{\ref{lesia}}
 \and A.~Bohn\inst{\ref{leiden}}
 \and M.-L.~Bolzer\inst{\ref{mpe}}
 \and M.~Bonnefoy\inst{\ref{ipag}}
 \and H.~Bonnet\inst{\ref{esog}}
 \and G.~Bourdarot\inst{\ref{mpe},\ref{ipag}}
 \and W.~Brandner\inst{\ref{mpia}}
 \and F.~Cantalloube\inst{\ref{lam}}
 \and P.~Caselli \inst{\ref{mpe}}
 \and B.~Charnay\inst{\ref{lesia}}
 \and G.~Chauvin\inst{\ref{ipag}}
 \and E.~Choquet\inst{\ref{lam}}
 \and V.~Christiaens\inst{\ref{monash}}
 \and Y.~Cl\'enet\inst{\ref{lesia}}
 \and V.~Coud\'e~du~Foresto\inst{\ref{lesia}}
 \and A.~Cridland\inst{\ref{leiden}}
 \and R.~Dembet\inst{\ref{esog}}
 \and J.~Dexter\inst{\ref{colorado}}
 \and P.~T.~de~Zeeuw\inst{\ref{leiden},\ref{mpe}}
 \and A.~Drescher\inst{\ref{mpe}}
 \and G.~Duvert\inst{\ref{ipag}}
 \and A.~Eckart\inst{\ref{cologne},\ref{bonn}}
 \and F.~Eisenhauer\inst{\ref{mpe}}
 \and F.~Gao\inst{\ref{hambourg}}
 \and P.~Garcia\inst{\ref{centra},\ref{porto}}
 \and R.~Garcia~Lopez\inst{\ref{dublin},\ref{mpia}}
 \and E.~Gendron\inst{\ref{lesia}}
 \and R.~Genzel\inst{\ref{mpe}}
 \and S.~Gillessen\inst{\ref{mpe}}
 \and J.~H.~Girard\inst{\ref{stsci}}
 \and X.~Haubois\inst{\ref{esoc}}
 \and G.~Hei\ss el\inst{\ref{lesia}}
 \and Th.~Henning\inst{\ref{mpia}}
 \and S.~Hinkley\inst{\ref{exeter}}
 \and S.~Hippler\inst{\ref{mpia}}
 \and M.~Horrobin\inst{\ref{cologne}}
 \and M.~Houll\'e\inst{\ref{lam}}
 \and Z.~Hubert\inst{\ref{ipag}}
 \and L.~Jocou\inst{\ref{ipag}}
 \and J.~Kammerer\inst{\ref{stsci}}
 \and M.~Keppler\inst{\ref{mpia}}
 \and P.~Kervella\inst{\ref{lesia}}
 \and L.~Kreidberg\inst{\ref{mpia}}
 \and V.~Lapeyr\`ere\inst{\ref{lesia}}
 \and J.-B.~Le~Bouquin\inst{\ref{ipag}}
 \and P.~L\'ena\inst{\ref{lesia}}
 \and D.~Lutz\inst{\ref{mpe}}
 \and A.-L.~Maire\inst{\ref{liege},\ref{mpia}}
 \and A.~M\'erand\inst{\ref{esog}}
 \and P.~Molli\`ere\inst{\ref{mpia}}
 \and J.~D.~Monnier\inst{\ref{umich}}
 \and D.~Mouillet\inst{\ref{ipag}}
 \and E.~Nasedkin\inst{\ref{mpia}}
 \and T.~Ott\inst{\ref{mpe}}
 \and G.~P.~P.~L.~Otten\inst{\ref{lam},\ref{asiaa}}
 \and C.~Paladini\inst{\ref{esoc}}
 \and T.~Paumard\inst{\ref{lesia}}
 \and K.~Perraut\inst{\ref{ipag}}
 \and G.~Perrin\inst{\ref{lesia}}
 \and O.~Pfuhl\inst{\ref{esog}}
 \and E.~Rickman\inst{\ref{esa}}
 \and L.~Pueyo\inst{\ref{stsci}}
 \and J.~Rameau\inst{\ref{ipag}}
 \and G.~Rousset\inst{\ref{lesia}}
 \and Z.~Rustamkulov \inst{\ref{stsci}}
 \and M.~Samland \inst{\ref{stockholm}}
 \and T.~Shimizu \inst{\ref{mpe}}
 \and D.~Sing \inst{\ref{stsci}}
 \and J.~Stadler\inst{\ref{mpe}}
 \and T.~Stolker\inst{\ref{leiden}}
 \and O.~Straub\inst{\ref{mpe}}
 \and C.~Straubmeier\inst{\ref{cologne}}
 \and E.~Sturm\inst{\ref{mpe}}
 \and L.~J.~Tacconi\inst{\ref{mpe}}
 \and E.F.~van~Dishoeck\inst{\ref{leiden},\ref{mpe}}
 \and A.~Vigan\inst{\ref{lam}}
 \and F.~Vincent\inst{\ref{lesia}}
 \and S.~D.~von~Fellenberg\inst{\ref{mpe}}
 \and K.~Ward-Duong\inst{\ref{amherst}}
 \and F.~Widmann\inst{\ref{mpe}}
 \and E.~Wieprecht\inst{\ref{mpe}}
 \and E.~Wiezorrek\inst{\ref{mpe}}
 \and J.~Woillez\inst{\ref{esog}}
 \and S.~Yazici\inst{\ref{mpe}}
 \and A.~Young\inst{\ref{mpe}}
 \and  the GRAVITY Collaboration}
\institute{
   LESIA, Observatoire de Paris, PSL, CNRS, Sorbonne Universit\'e, Universit\'e de Paris, 5 place Janssen, 92195 Meudon, France
\label{lesia}      \and
   European Southern Observatory, Karl-Schwarzschild-Stra\ss e 2, 85748 Garching, Germany
\label{esog}      \and
   Department of Astronomy, California Institute of Technology, Pasadena, CA 91125, USA
\label{caltech}      \and
   Center for Astrophysics and Planetary Science, Department of Astronomy, Cornell University, Ithaca, NY 14853, USA
\label{cornell}      \and
   Institute of Astronomy, University of Cambridge, Madingley Road, Cambridge CB3 0HA, United Kingdom
\label{cam}      \and
   Max Planck Institute for extraterrestrial Physics, Giessenbachstra\ss e~1, 85748 Garching, Germany
\label{mpe}      \and
   Universit\'e Grenoble Alpes, CNRS, IPAG, 38000 Grenoble, France
\label{ipag}      \and
   Universidade de Lisboa - Faculdade de Ci\^encias, Campo Grande, 1749-016 Lisboa, Portugal
\label{lisboa}      \and
   CENTRA - Centro de Astrof\' isica e Gravita\c c\~ao, IST, Universidade de Lisboa, 1049-001 Lisboa, Portugal
\label{centra}      \and
   Max Planck Institute for Astronomy, K\"onigstuhl 17, 69117 Heidelberg, Germany
\label{mpia}      \and
   Leiden Observatory, Leiden University, P.O. Box 9513, 2300 RA Leiden, The Netherlands
\label{leiden}      \and
   Aix Marseille Univ, CNRS, CNES, LAM, Marseille, France
\label{lam}      \and
   School of Physics and Astronomy, Monash University, Clayton, VIC 3800, Melbourne, Australia
\label{monash}      \and
  JILA and Department of Astrophysical and Planetary Sciences, University of Colorado, Boulder, CO 80309, USA
\label{colorado}      \and
   1. Institute of Physics, University of Cologne, Z\"ulpicher Stra\ss e 77, 50937 Cologne, Germany
\label{cologne}      \and
   Max Planck Institute for Radio Astronomy, Auf dem H\"ugel 69, 53121 Bonn, Germany
\label{bonn}      \and
   Hamburger Sternwarte, Universit\"at Hamburg, Gojenbergsweg 112, 21029 Hamburg, Germany
\label{hambourg}      \and
   Universidade do Porto, Faculdade de Engenharia, Rua Dr. Roberto Frias, 4200-465 Porto, Portugal
\label{porto}      \and
   School of Physics, University College Dublin, Belfield, Dublin 4, Ireland
\label{dublin}    \and
   Space Telescope Science Institute, Baltimore, MD 21218, USA
\label{stsci}   \and
   European Southern Observatory, Casilla 19001, Santiago 19, Chile
\label{esoc}      \and
   University of Exeter, Physics Building, Stocker Road, Exeter EX4 4QL, United Kingdom
\label{exeter}     \and
   STAR Institute, Universit\'e de Li\`ege, All\'ee du Six Ao\^ut 19c, B-4000 Li\`ege, Belgium
\label{liege}      \and
   Astronomy Department, University of Michigan, Ann Arbor, MI 48109 USA
\label{umich}       \and
   Academia Sinica Institute of Astronomy and Astrophysics, 11F Astronomy-Mathematics Building, NTU/AS campus, No. 1, Section 4, Roosevelt Rd., Taipei 10617, Taiwan
\label{asiaa}      \and
   European Space Agency (ESA), ESA Office, Space Telescope Science Institute, Baltimore, MD 21218, USA
\label{esa}        \and
   Department of Astronomy, Stockholm University, Stockholm, Sweden
\label{stockholm}   \and
   Five College Astronomy Department, Amherst College, Amherst, MA 01002, USA
\label{amherst}      \and
   Research School of Astronomy \& Astrophysics, Australian National University, ACT 2611, Australia
\label{australie}
}



  \abstract
   {}
   {
 We aim to demonstrate that the presence and mass of an exoplanet can now be effectively derived from the astrometry of another exoplanet.
   }
   {We combined previous astrometry of $\beta$ Pictoris b with a new set of observations from the GRAVITY interferometer. The orbital motion of $\beta$ Pictoris b is fit using Markov chain Monte Carlo simulations in Jacobi coordinates. 
   The inner planet, $\beta$ Pictoris c, was also reobserved at a separation of 96\,mas, confirming the previous orbital estimations.}
   {
   From the astrometry of planet b only, we can (i) detect the presence of $\beta$ Pictoris c and (ii) constrain its mass to  $10.04^{+4.53}_{-3.10}\,M_{\rm Jup}$. If one adds the astrometry of $\beta$ Pictoris c, the mass is narrowed down to $9.15^{+1.08}_{-1.06}\,M_{\rm Jup}$.
   The inclusion of radial velocity measurements does not affect the orbital parameters significantly, but it does slightly  decrease the mass estimate to $8.89^{+0.75}_{-0.75}\,M_{\rm Jup}$.
With a semimajor axis of $2.68\pm0.02$\,au, a period of $1221\pm15$ days, and an eccentricity of  $0.32\pm0.02$,  the orbital parameters of $\beta$ Pictoris c are now constrained as precisely as those of $\beta$ Pictoris b.
 The orbital configuration is compatible with a high-order mean-motion resonance (7:1). The impact of the resonance on the planets' dynamics would then be negligible with respect to the secular perturbations, which might have played an important role in the eccentricity excitation of the outer planet.
  }
 {}


   \maketitle
%

\section{Introduction}

The formation and evolution of giant exoplanets is an intense field of research. Several formation scenarios, ranging from gravitational instabilities \citep[e.g.,][]{bossFormationGiantPlanets2011,nayakshinDawesReviewTidal2017} to a variety of core-accretion models
\citep[e.g.,][]{alibertModelsGiantPlanet2005,emsenhuberNewGenerationPlanetary2020a}, are still actively debated. The issue is the lack of observables with which to distinguish between the different scenarios. One solution is to analyze the atmospheric composition and search for formation signatures  \citep{obergEffectsSnowlinesPlanetary2011,mordasiniImprintExoplanetFormation2016,madhusudhanAtmosphericSignaturesGiant2017}. Another solution consists in measuring the energy dissipation during formation as a function of the final exoplanetary mass \citep{2013A&A...558A.113M,marleauConstrainingInitialEntropy2014a}.

This paper focuses on the latter. Obtaining a dynamical mass for young directly imaged exoplanets is difficult. Only a handful of these objects have published dynamical masses: $\beta$ Pictoris\,b \citep{2018NatAs...2..883S,2019ApJ...871L...4D,lagrangeUnveilingPictorisSystem2020,vandalDynamicalMassEstimates2020}, $\beta$ Pictoris\,c \citep{nowakDirectConfirmationRadialvelocity2020}, PDS\ 70\,b \citep{wangConstrainingNaturePDS2021}, and HR8799\,e
\citep{2021ApJ...915L..16B}.
One of the main difficulties of a direct measurement is the long orbital period of directly imaged exoplanets. Another is the fact that young stars are often pulsating (for example, $\beta$ Pictoris is a $\delta$ Scuti variable), which makes accurate radial velocity (RV) measurements difficult.

An efficient technique for measuring dynamical masses is observing multi-planetary systems and detecting the mutual influence of planets. The main historical example is the prediction of Neptune from the irregularities in the orbit of Uranus by \citet{LeVerrier:1846}. A more recent example is the
 first exoplanetary system detected around PSR B1257+12  \citep{1992Natur.355..145W}, where the mutual interactions of the planets were used to confirm their masses \citep[e.g.,][]{2003ApJ...591L.147K}.
Last, transit timing variations \citep{agolDetectingTerrestrialPlanets2005} has become a key technique for obtaining the mass of transiting planets \citep[e.g.,][]{2020A&A...642A..49D}. In this paper we demonstrate that optical interferometry is now able to detect and measure the mass of an exoplanet solely from the astrometry of another exoplanet at larger separation.

We focus on the $\beta$ Pictoris system, where both the outer planet \citep[b;][]{lagrangeGiantPlanetImaged2010a} and the inner planet \citep[c;][]{lagrangeEvidenceAdditionalPlanet2019} have well-characterized orbital parameters \citep{lagrangeUnveilingPictorisSystem2020,nowakDirectConfirmationRadialvelocity2020}.
We use the GRAVITY instrument  \citep{gravitycollaborationFirstLightGRAVITY2017}, a near-infrared interferometer operating at the Very Large Telescope (VLT) at Cerro Paranal. The interferometer has been designed to
 theoretically reach 10\,$\mu$as accuracy \citep{lacourReachingMicroarcsecondAstrometry2014} and has effectively demonstrated this level of accuracy on SgrA* flares  at the center of the Milky Way \citep{gravitycollaborationDetectionOrbitalMotions2018}. On exoplanets, it has demonstrated a typical 50\,$\mu$as accuracy \citep{gravitycollaborationFirstDirectDetection2019,lagrangeUnveilingPictorisSystem2020}.

 In Sect.~\ref{sec:obs} we present the new data reduction and additional, recent astrometric observations. Section~\ref{sec:jac} presents the restricted three-body model that we use for the Markov chain Monte Carlo (MCMC) fits. In Sect.~\ref{sec:ind} $\beta$ Pictoris c is detected from $\beta$ Pictoris b astrometry only. In Sect.~\ref{sec:full} we perform a fit that includes $\beta$ Pictoris c astrometry and RVs. In Sect.~\ref{ref:dis} we discuss dynamical insights from the orbital solutions.
Section~\ref{sec:conc} is the conclusion.


\section{Observations}
\label{sec:obs}

We obtained three additional observations of the $\beta$ Pictoris system with GRAVITY. Beta Pictoris c was observed during the night of 6 January 2021 and planet b during the nights of 7 January and 27 August 2021.
The weather conditions ranged from very good to bellow average (on 27 January).  The log of the observations are presented in Table~\ref{tab:log}. These observations were part of the ExoGRAVITY large program \citep{2020SPIE11446E..0OL}, taken with a similar observation sequence as for past exoplanet detections \citep{gravitycollaborationFirstDirectDetection2019,gravitycollaborationPeeringFormationHistory2020}: The fringe tracker
\citep{lacourGRAVITYFringeTracker2019} observes the star while the science camera \citep{2014SPIE.9146E..29S} alternately observes the star and the exoplanet.

The new observations of $\beta$ Pictoris b and c were processed with the Public Release 1.5.0  (1 July 2021\footnote{\url{https://www.eso.org/sci/software/pipelines/gravity/}}) of the ESO GRAVITY pipeline \citep{2014SPIE.9146E..2DL}. This new version has slightly better astrometric capabilities by accounting for the effect of differential astigmatism \citep{gravitycollaborationImprovedGRAVITYAstrometric2021}. We also reprocessed previously published GRAVITY data sets with the same pipeline version.
 From the seven epochs published by \citet{lagrangeUnveilingPictorisSystem2020}, we discarded two data sets: one from 2 November 2019 and one taken on 7 January 2020. In both circumstances, the very low coherence time ($\tau_0<1.5\,$ms) and limited number of exposures made the calculation of astrometric uncertainties difficult.
  The updated and new astrometric values are presented in Table~\ref{tb:astrometry}.

\begin{table}
  \begin{center}
    \begin{tabular}{c c c c c c}
      \hline
      \hline
       MJD & $\Delta\RA$ & $\Delta\DEC$ & $\sigma_{\Delta\RA}$ & $\sigma_{\Delta\DEC}$ & $\rho$ \\
      (days) & (mas) & (mas) & (mas) & (mas) & - \\
      \hline
      \multicolumn{6}{c}{ $\beta$ Pictoris b} \\
      \hline
      58383.378 & 68.47 & 126.38 &0.05 & 0.07 & -0.86\\
      58796.170 &145.51 & 248.59 &0.11 & 0.05 & -0.85\\
      58798.356 & 145.65  & 249.21 & 0.03 & 0.09 & -0.44\\
      58855.065 & 155.41 & 264.33 & 0.15 & 0.29 & -0.71\\
      58889.139 &160.96 & 273.41 & 0.06 & 0.13& -0.56\\
      59221.238& 211.59 & 352.62&0.02 & 0.05 & -0.10\\
      59453.395& 240.63 & 397.89 & 0.09 & 0.04 & -0.91 \\
       \hline
      \multicolumn{6}{c}{ $\beta$ Pictoris c} \\
      \hline
      58889.140 & -67.36 & -112.59 & 0.17 & 0.24 & -0.80 \\
      58891.065 & -67.67 & -113.20 & 0.11 & 0.19 & -0.54 \\
      58916.043 & -71.88 & -119.60 & 0.07 & 0.14 & -0.52 \\
      59220.163 & --52.00 & -80.86 & 0.20 & 0.34 & -0.26 \\
      \hline
      \hline
    \end{tabular}
    \caption{Relative astrometry of $\beta$ Pictoris c extracted from  VLTI/GRAVITY observations. The Pearson coefficient ($\rho$) quantifies the correlation between the RA and Dec uncertainties.}
    \label{tb:astrometry}
  \end{center}
\end{table}

\section{A restricted three-body problem for multi-planetary systems}
\label{sec:jac}

The Newtonian three-body problem is often addressed in a restricted version, where the mass of one body is negligible with respect to the other two (e.g., a satellite within the gravitational field of Earth and the Sun). Our case is different because the masses of the exoplanets are similar. However, a different restricted version of the three-body problem can be derived for systems with comparable masses. It uses Jacobi coordinates \citep{plummerIntroductoryTreatiseDynamical1918}.

\citet{beustSymplecticIntegrationHierarchical2003} already developed the formalism for any $N$-body system consisting of hierarchically nested orbits. Here we applied it to the specific case of a solar-mass object with two planets. The Lagrangian writes $L=T-V$, where T is the kinetic energy and V the potential energy. With three bodies interacting by gravitational force, this writes ($G$ is the constant of gravitation):
\begin{equation}
L= \frac{1}{2} (m_\star \vec{\dot x_\star}^2+m_b \vec{\dot x_b}^2+m_c \vec{\dot x_c}^2) + \frac{G m_\star  m_b  }{|\vec {x_b}-\vec {x_\star}|}+ \frac{G m_\star m_c}{|\vec {x_c}-\vec {x_\star}|}+ \frac{G m_b m_c }{|\vec {x_c}-\vec {x_b}|}
\label{eq:lagrange}
,\end{equation}
where $m_\star$ and $\vec {x_\star}$ correspond to the mass and position of the star, and $m_b$, $\vec {x_b}$, $m_c$, and $\vec {x_c}$, the mass and position of exoplanets b and c, respectively.
To transform these equations into Jacobi coordinates, we set, according to  \citet{wisdomSymplecticMapsNbody1991}:
\begin{eqnarray}
\vec q &=& \vec {x_c}-\vec {x_\star}\\
\vec Q &=& \vec {x_b}-(m_\star{}\vec{x_\star{}}+m_c \vec{x_c})\nu^{-1}\\
\vec R &=& (m_\star{}\vec{x_\star{}} + m_b \vec{x_b} + m_c \vec{x_c}) M^{-1}
,\end{eqnarray}
where $\vec q$ represents the position of planet c relative to the star, $\vec Q$ the position of planet b relative to the center of mass of the system (star, planet c), and $\vec{R}$ the position of the center of mass of the system. We also set $M= m_\star+m_b  +m_c$ and $\nu = m_\star+m_c$.

The vectors $\vec{x_{\star{}, b, c}}$ can be rewritten in terms of these Jacobi coordinates:
\begin{eqnarray}
\vec{x_\star} &=& \vec R - m_b M^{-1} \vec Q - m_c\nu^{-1}\vec q\\
\vec{x_b} &=& \vec{R} + (m_{\star}+m_b)M^{-1}\vec{Q}\\
\vec{x_c} &=& \vec R - m_bM^{-1}\vec Q + m_\star \nu^{-1}\vec q
.\end{eqnarray}

In the barycentric reference frame, where the velocity of the center of mass $\dot{\vec R}$ is zero, we have:
\begin{eqnarray}
\dot{\vec{x_\star}} &=& - m_b M^{-1}\dot{\vec Q} - m_c\nu^{-1}\dot{\vec q}\\
\dot{\vec{x_b}} &=& (m_{\star}+m_b )M^{-1}\dot{\vec{Q}}\\
\dot{\vec{x_c}} &=& - m_b M^{-1}\dot{\vec Q} + m_\star \nu^{-1}\dot{\vec q}\\
\vec{x_b} - \vec{x_\star} &=& \vec{Q} + m_c \nu^{-1} \vec{q} \\
\vec{x_c} - \vec{x_\star} &=& \vec{q} \\
\vec{x_c} - \vec{x_b} &=& \vec{Q} - m_\star \nu^{-1}\vec{q}
.\end{eqnarray}

The terms of Eq.~(\ref{eq:lagrange}) are replaced with these new expressions:
\begin{eqnarray}
T &=& {1}/{2}\left(m_c m_\star \nu^{-1}\vec{\dot q}^2 + m_b\nu M^{-1} \vec{\dot Q}^2 \right) \\
V &=& -\frac{G m_\star m_b}{\left|\vec{Q}+m_c\nu^{-1}\vec{q}\right|}
- \frac{G m_\star m_c}{\left|\vec{q}\right|}
- \frac{G m_b m_c}{\left|\vec{Q}-m_\star\nu^{-1}\vec{q}\right|}
.\end{eqnarray}

In the restricted case that matters to us ($|\vec {x_b}- \vec {x_\star}| >> |\vec {x_c}- \vec {x_\star}|$, i.e. $|\vec{Q}|>>|\vec{q}|$), the potential energy becomes, to first order:
\begin{equation}
V = - \frac{G m_\star m_c}{\left|\vec{q}\right|} -\frac{G m_b\nu}{\left|\vec{Q}\right|}
,\end{equation}
and the Lagrangian can be approximated by:
\begin{equation}
L= \frac{1}{2} \left( m_c  m_\star\nu^{-1} \vec{\dot q}^2
+m_b\nu M^{-1} \vec{\dot Q}^2\right) + \frac{G m_\star  m_c  }{|\vec q|}+ \frac{G m_b\nu}{|\vec{Q}|} \label{eq:redL}
.\end{equation}

In this expression, the two Jacobi variables $\vec Q$ and $\vec q$ are decoupled, and the Lagrangian can be written as the sum of two terms, $L=L_q+L_Q$, with:
\begin{eqnarray}
L_q &=&\frac{1}{2}\frac{m_c  m_\star}{m_c+m_\star} \vec{\dot q}^2+\frac{G m_\star  m_c  }{|\vec q|} \\
L_Q &=& \frac{1}{2}\frac{m_b(m_\star+m_c)}{m_\star+m_b+m_c} \vec{\dot Q}^2 +\frac{G m_b(m_\star+m_c)}{|\vec{Q}|}
.\end{eqnarray}

 The term $L_q$ corresponds to the Lagrangian of the classical two-body problem that describes the orbit of planet c around the star, whereas $L_Q$ is the Lagrangian of the two-body problem corresponding to planet b orbiting around  a virtual particle of mass $m_c+m_\star$ located at the center of mass of  the system (star, planet c). Both quantities can be solved analytically. This is how we modeled the orbital motion of $\beta$ Pictoris b and c. The model should be the same as that used in \citet{Brandt2021betapic}. We validate that our orbital model is sufficiently accurate for our GRAVITY astrometry in Appendix~\ref{sec:nbody}.

    \begin{figure}
      \resizebox{\hsize}{!}{\includegraphics{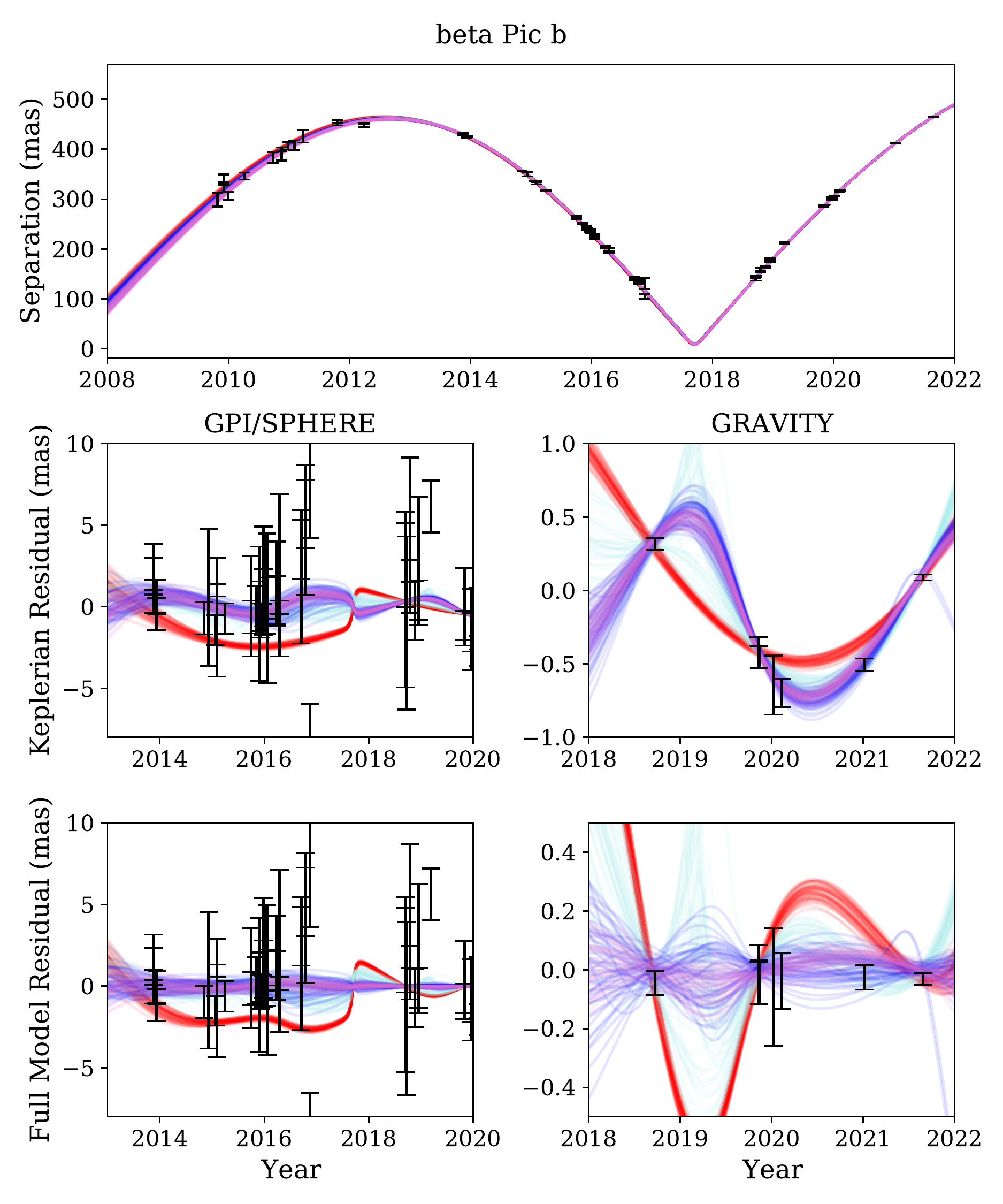}}
      \caption{Projected separation of $\beta$ Pictoris b as a function of time.\ The one-planet and two-planet\ model fits that use only the b astrometry are shown in red and cyan, respectively. These models are described in Sect.~\ref{sec:ind}.
      The two-planet model fit that uses b and c astrometry is also plotted, in blue, and the fit that also uses the RV is plotted in purple (Sect.~\ref{sec:full}). 
      The top row shows the orbit models and all of the data. 
      The middle row shows the residuals after subtracting a pure Keplerian orbit for planet b based on the orbital parameters from the two-planet model using b and c astrometry. The bottom row shows the residual after accounting for the perturbation of planet c. We note that, although the red one-planet\ model is a pure Keplerian orbit, it is not a flat line in the middle row because the best-fit one-planet Keplerian model also attempts to fit the perturbations due to the second planet.  }
      \label{fig:betaB}
    \end{figure}

\section{Detection of $\beta$ Pic c ``with the point of [a] pen''}
\label{sec:ind}

 Here,  instead of a pen\footnote{\citet{Arago:1846} famously referred to Le Verrier's theoretical prediction of Neptune's existence as a discovery made with the point of his pen.}, we use the \texttt{orbitize!} code\footnote{\url{https://orbitize.readthedocs.io}} \citep{Blunt2020}.
In this section we fit the relative astrometry of $\beta$ Pictoris b only to assess whether we can indirectly detect $\beta$ Pictoris c. We fit both a one-planet model and a two-planet model to only the astrometry of $\beta$ Pictoris b. The one-planet model is a repeat of the fit done in \citet{gravitycollaborationPeeringFormationHistory2020}, using the same eight orbital parameters: semimajor axis ($a_b$), eccentricity ($e_b$), inclination ($i_b$), argument of periastron ($\omega_b$), position angle of the ascending node ($\Omega_b$), epoch of periastron in fractional orbital periods after MJD 59,000 ($\tau_b$), system parallax, and total mass ($M_\textrm{tot}$). We used all the same priors as \citet{gravitycollaborationPeeringFormationHistory2020}. The two-planet model  fit adds orbital elements for a second planet ($a_c$, $e_c$, $i_c$, $\omega_c$, $\Omega_c$, $\tau_c$) as well as replacing total mass with component masses ($M_*$ for the star and $M_b$ and $M_c$ for planets b and c, respectively). The priors on most of the orbital elements of $\beta$ Pictoris c are the same as for b, except for a log uniform prior from 0.1 to 9 au for $a_c$. We used the same prior on $M_\textrm{tot}$ as $M_*$. We used a log uniform prior of 1 to 50 $M_\textrm{Jup}$ for $M_c$. We fixed the mass of $M_b$ to 10 $M_\textrm{Jup}$ since our orbital model described in Sect. \ref{sec:jac} cannot particularly constrain $M_b$ unless $M_*$ is known to $<$ 1\% precision.

In both cases we used the parallel-tempered affine-invariant sampler in \texttt{ptemcee} \citep{foreman-mackeyEmceeMCMCHammer2013, vousdenDynamicTemperatureSelection2016}, using 20 temperatures and 1000 walkers per temperature. We obtained 30,000 samples of the posterior per walker after a ``burn-in'' of 10,000 steps for each walker in the one-planet model fit. In the two-planet model  fit, we obtained 5,000 samples of the posterior from each walker after a burn-in of 55,000 steps for each walker.  The posteriors for the parameters are given in  Table~\ref{tab:allResultsl}. For the one-planet fit, there are no assumptions -- and therefore no constraints -- on planet c. The two-planet fit is able to indirectly measure a distinct mass and $a_c$ for the second planet.

To assess whether adding a second planet significantly improves the fit to the data, we computed the Bayesian information criterion (BIC) of the maximum likelihood model for both models. The one-planet model gives a BIC of 1247, while the two-planet model fit gives a BIC of 1784. The difference of 537 in the BIC indicates definitively that we have indirectly detected $\beta$ Pictoris c using only the relative astrometry of $\beta$ Pictoris b. For comparison, BIC changes between 10 and 100 have been used to show significant detections of outer planets in RV data \citep{Christiansen2017,Bryan2019}.
This is also reflected in the residuals to the orbit fits shown in Fig. \ref{fig:betaB}. In both the residuals to the GPI/SPHERE astrometry and the residual to the GRAVITY astrometry, the one-planet model fit in red fails to fully fit all of the data points and is systematically off. This corroborates the large change in BIC going from the one-planet to the two-planet model.

In addition to the detection of $\beta$ Pictoris c, we also determined its mass, obtaining $10.04^{+4.53}_{-3.10}\,M_{\rm Jup}$. The  estimate is slightly high, although it is still consistent with previous mass derivations \citep{lagrangeEvidenceAdditionalPlanet2019,vandalDynamicalMassEstimates2020,Brandt2021betapic}.
The eccentricity was also determined, but with large uncertainties: $e_c=0.90^{+0.08}_{-0.26}$.
The value is high, although the 95\% credible interval (in Table~\ref{tab:allResultsl}) is still consistent with previously published estimations. However, we have not included all of the data, and this merely demonstrates the ability of GRAVITY to indirectly detect a second planet in the system. The direct detection of $\beta$ Pictoris c brings much better orbital constraints, as shown in the following section.

    \begin{figure}
      \resizebox{\hsize}{!}{\includegraphics{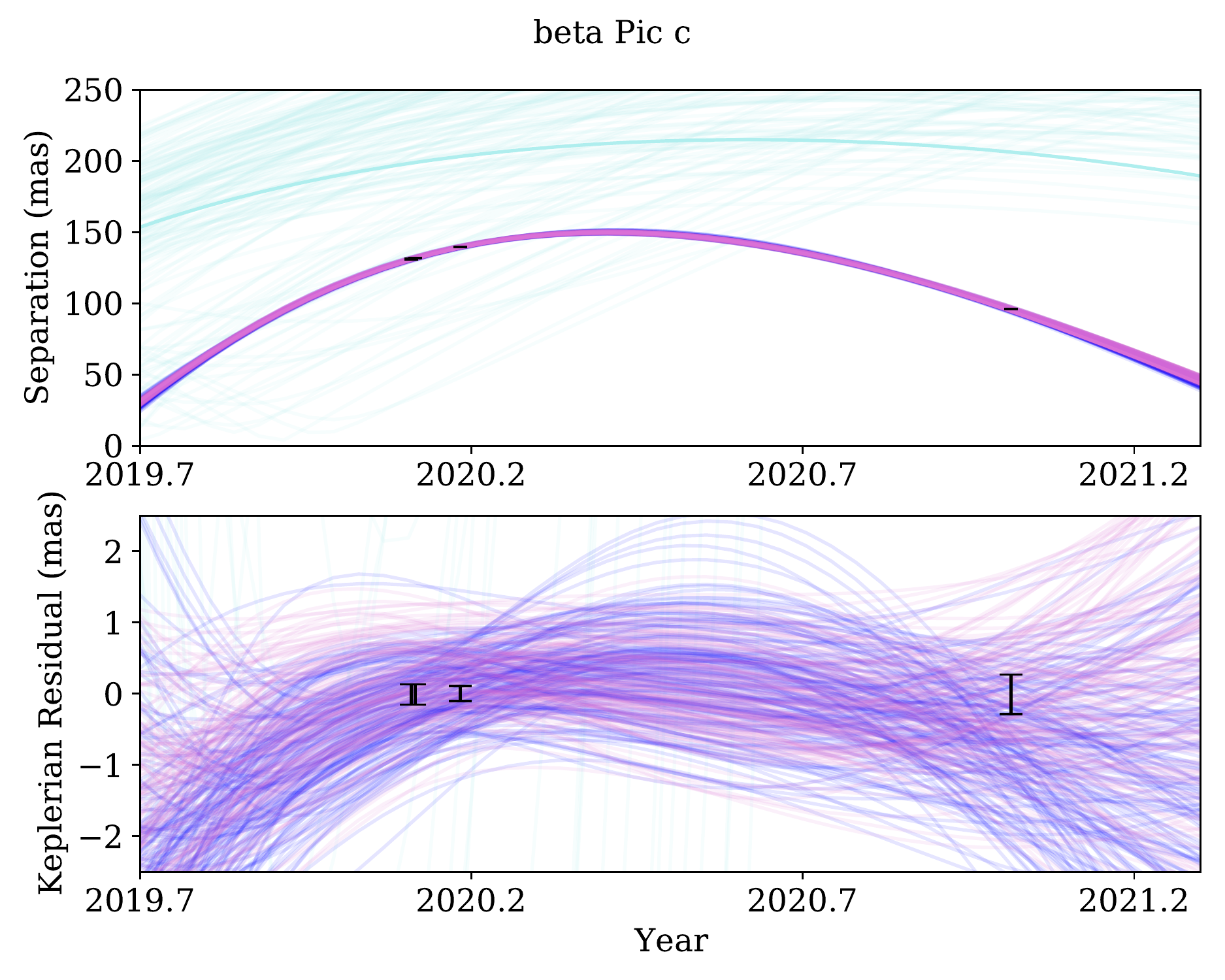}}
      \caption{Projected separation of $\beta$  Pictoris c as a function of time. The top row shows the data (black error bars) and models (same coloring as Fig.~\ref{fig:betaB}). The bottom row shows the same data but after subtraction of the median orbit of the two-planet model fit from b and c astrometry. The two-planet model  that uses only b astrometry (in teal) only weakly constraints the angular separation of planet c, and a direct detection of planet c is needed to obtain precise orbital constraints.}
      \label{fig:betaC}
    \end{figure}

\section{Refined masses and orbits of $\beta$ Pic b and c}
\label{sec:full}

More accurate orbital parameters of $\beta$ Pictoris c can be derived if we include the relative astrometry of c. This, in turn, gives a better mass estimation of c. The two-planet model fit was repeated, but with the addition of the $\beta$ Pictoris c GRAVITY astrometry. The same MCMC walker parameters were used to obtain the 500,000 samples of the posterior. The posteriors are plotted in blue in both Fig.~\ref{fig:betaB} ($\beta$ Pictoris b) and Fig.~\ref{fig:betaC} ($\beta$ Pictoris c). The new  $\beta$ Pictoris c observation now accurately constrains the eccentricity of the orbit ($e=0.32^{+0.03}_{-0.02}$) and the mass of $\beta$ Pictoris c ($9.15^{+1.08}_{-1.06}\,M_{\rm Jup}$). The mass of $\beta$ Pictoris c is consistent and comparable in precision to the mass obtained from stellar RVs and absolute astrometry \citep{lagrangeEvidenceAdditionalPlanet2019,vandalDynamicalMassEstimates2020,Brandt2021betapic}. This improves the robustness of the mass estimate since GRAVITY astrometry is not affected by systematic errors in stellar RVs on young stars or in absolute astrometry on bright stars.

Last, we performed a two-planet fit that included the RV data of the star from \citet{lagrangeUnveilingPictorisSystem2020} that was reprocessed by \citet{vandalDynamicalMassEstimates2020}. The inclusion of the RV data allowed us to fit for the mass of $\beta$ Pictoris b, which we had previously left fixed. We used a log uniform prior between 1 and 50 $M_\textrm{Jup}$ for its mass prior. For the MCMC analysis, we used two temperatures and 1000 walkers per temperature. Each walker underwent a 10,000-step burn-in before 1,000 samples of the posterior were obtained from each walker.
We find a semimajor axis of $2.68\pm0.02$\,au, corresponding to an orbital period of $1221\pm15\ $days.
The mass estimate of $\beta$~Pictoris~c is still within the previous fit uncertainties, at $8.89^{+0.75}_{-0.75}\,M_{\rm Jup}$, and we estimate a mass of $11.90^{+2.93}_{-3.04}\,M_{\rm Jup}$ for $\beta$ Pictoris b. Both component masses agree well  with the latest dynamical mass estimates \citep{vandalDynamicalMassEstimates2020,Brandt2021betapic}. The two masses are also in very good agreement with a hot-core accretion scenario \citep{2013A&A...558A.113M}, as proposed by \citet{nowakDirectConfirmationRadialvelocity2020}.

\begin{table*}
    \centering
    \begin{tabular}{cccccc}
        \hline
        \hline
         &  & one-planet model & two-planet model& two-planet model& two-planet model \\
        \hline
        Parameter & Unit & b astrometry &b astrometry & b and c astrometry & RV + b and c astrometry \\
        \hline
        $a_b$ & au & $10.01^{+0.02(+0.05)}_{-0.03(-0.05)}$ & $10.01^{+0.03(+0.06)}_{-0.03(-0.06)}$ & $9.95^{+0.03(+0.06)}_{-0.02(-0.05)}$ & $9.93^{+0.03(+0.05)}_{-0.03(-0.05)}$ \\
        $e_b$ & & $0.104^{+0.003(+0.006)}_{-0.003(-0.006)}$ & $0.110^{+0.004(+0.008)}_{-0.003(-0.007)}$ & $0.106^{+0.004(+0.007)}_{-0.004(-0.007)}$ & $0.103^{+0.003(+0.006)}_{-0.003(-0.005)}$ \\
        $i_b$ & deg & $89.01^{+0.01(+0.02)}_{-0.01(-0.01)}$ & $89.00^{+0.02(+0.04)}_{-0.04(-0.10)}$ & $88.99^{+0.01(+0.02)}_{-0.01(-0.02)}$ & $89.00^{+0.00(+0.01)}_{-0.01(-0.02)}$ \\
        $\omega_b$ & deg & $198.4^{+2.9(+5.5)}_{-3.1(-6.2)}$ & $199.7^{+3.2(+6.3)}_{-3.5(-7.4)}$ & $203.2^{+2.8(+5.2)}_{-3.2(-7.0)}$ & $199.3^{+2.8(+4.9)}_{-3.1(-6.1)}$ \\
        $\Omega_b$ & deg & $31.79^{+0.01(+0.01)}_{-0.01(-0.02)}$ & $31.90^{+0.03(+0.06)}_{-0.02(-0.05)}$ & $31.80^{+0.00(+0.01)}_{-0.01(-0.02)}$ & $31.79^{+0.01(+0.02)}_{+0.00(-0.01)}$ \\
        $\tau_b$ & & $0.717^{+0.009(+0.018)}_{-0.009(-0.019)}$ & $0.723^{+0.010(+0.019)}_{-0.011(-0.022)}$ & $0.730^{+0.009(+0.017)}_{-0.010(-0.021)}$ & $0.719^{+0.008(+0.014)}_{-0.010(-0.019)}$ \\
        \hline
        $a_c$ & au & & $4.21^{+1.38(+3.43)}_{-1.33(-1.48)}$ & $2.61^{+0.06(+0.12)}_{-0.06(-0.11)}$ & $2.68^{+0.02(+0.05)}_{-0.02(-0.04)}$ \\
        $e_c$ & & & $0.90^{+0.08(+0.10)}_{-0.26(-0.71)}$ & $0.32^{+0.03(+0.06)}_{-0.02(-0.05)}$ & $0.32^{+0.02(+0.03)}_{-0.02(-0.04)}$ \\
        $i_c$ & deg & & $90.52^{+10.81(+25.59)}_{-9.01(-26.24)}$ & $88.92^{+0.10(+0.20)}_{-0.10(-0.20)}$ & $88.95^{+0.09(+0.19)}_{-0.10(-0.20)}$ \\
        $\omega_c$ & deg & & $27.1^{+20.4(+49.4)}_{-17.9(-25.7)}$ & $60.8^{+4.2(+9.1)}_{-3.7(-7.0)}$ & $66.0^{+1.8(+3.7)}_{-1.7(-3.7)}$ \\
        $\Omega_c$ & deg & & $55.70^{+7.08(+15.14)}_{-6.74(-14.47)}$ & $31.07^{+0.05(+0.10)}_{-0.04(-0.09)}$ & $31.06^{+0.04(+0.08)}_{-0.04(-0.08)}$ \\
        $\tau_c$ & & & $0.804^{+0.080(+0.132)}_{-0.143(-0.189)}$ & $0.708^{+0.012(+0.025)}_{-0.012(-0.023)}$ & $0.724^{+0.006(+0.012)}_{-0.006(-0.013)}$ \\
        \hline
        Parallax & mas & $51.44^{+0.12(+0.23)}_{-0.12(-0.24)}$ & $51.44^{+0.12(+0.23)}_{-0.12(-0.24)}$ & $51.44^{+0.12(+0.23)}_{-0.12(-0.24)}$ & $51.44^{+0.12(+0.24)}_{-0.12(-0.24)}$ \\
        \hline
        $M_*$ & $M_\odot$ & & $1.75^{+0.02(+0.05)}_{-0.03(-0.05)}$ & $1.73^{+0.03(+0.05)}_{-0.02(-0.04)}$ & $1.75^{+0.03(+0.05)}_{-0.02(-0.04)}$ \\
        $M_b$ & $M_\textrm{Jup}$ & & $10.00$ & $10.00$ & $11.90^{+2.93(+5.68)}_{-3.04(-6.18)}$ \\
        $M_c$ & $M_\textrm{Jup}$ & & $10.04^{+4.53(+8.66)}_{-3.10(-4.32)}$ & $9.15^{+1.08(+2.18)}_{-1.06(-2.11)}$ & $8.89^{+0.75(+1.49)}_{-0.75(-1.47)}$ \\
      \hline
      \hline
    \end{tabular}
    \caption{Posteriors for the main parameters of the orbital fits. 
    Orbital parameters are in Jacobi coordinates and follow the definitions in \citet{Blunt2020}, with the exception of the reference epoch for $\tau$ as MJD 59,000 (31 May 2020). For each orbital parameter, the median is reported along with the 68\% credible interval in super- and subscript (the 95\% credible interval is reported in parentheses). }
    \label{tab:allResultsl}
\end{table*}

\section{Stability, resonance, and evaporating bodies}
\label{ref:dis}

The $\beta$ Pictoris system has only recently joined the selective group of multi-planet directly imaged systems, along with HR 8799 \citep{maroisDirectImagingMultiple2008,maroisImagesFourthPlanet2010}, PDS 70 \citep{kepplerDiscoveryPlanetarymassCompanion2018,mullerOrbitalAtmosphericCharacterization2018,haffertTwoAccretingProtoplanets2019}, and TYC 8998-760-1 \citep{2020ApJ...898L..16B}. Because of the lack of planet detection and the struggle to constrain long-period orbits, the formation and dynamical evolution of cold giant planets remain largely unconstrained.

The eccentricity is thought to trace the formation and dynamical evolution of wide substellar objects. Indeed, wide-orbit giant planets and brown dwarfs appear to have different eccentricity distributions \citep{bowlerPopulationlevelEccentricityDistributions2020}. Low eccentricities and nearly planar planetary configurations are nevertheless usually associated with stable systems that formed via core accretion, such as the Solar System, even if post-formation giant impacts, planet-planet interactions, or scattering can excite orbital eccentricities afterward. With their orbits currently separated by $\sim 5$ relative Hill radii, the current configuration of the two planets is likely stable, a statement confirmed by Beust et al. (2021, in preparation). However, both planets presumably induce significant secular perturbations on each other due to their high masses. The eccentricity variations caused by these
perturbations are detailed in Appendix~\ref{sec:secular}. It is interesting to note that, in most solutions from the orbital fit, planet b periodically reduces its eccentricity to a negligible value, while the eccentricity
of planet c remains greater than $0.2$. This suggests that, contrary to planet c, the eccentricity of planet b may not be primordial, but rather the result of secular perturbations from planet c. Secular interaction also applies to inclination. It has long been noted that the midplane of the inner disk differs from that of the outer  \citep{1997MNRAS.292..896M,2000ApJ...539..435H,2001A&A...370..447A}  by a few degrees. Secular perturbations arising from both planets themselves having inclination oscillations could indeed cause this phenomenon.

\begin{figure}[h]
    \centering
    \includegraphics[width=0.75\linewidth]{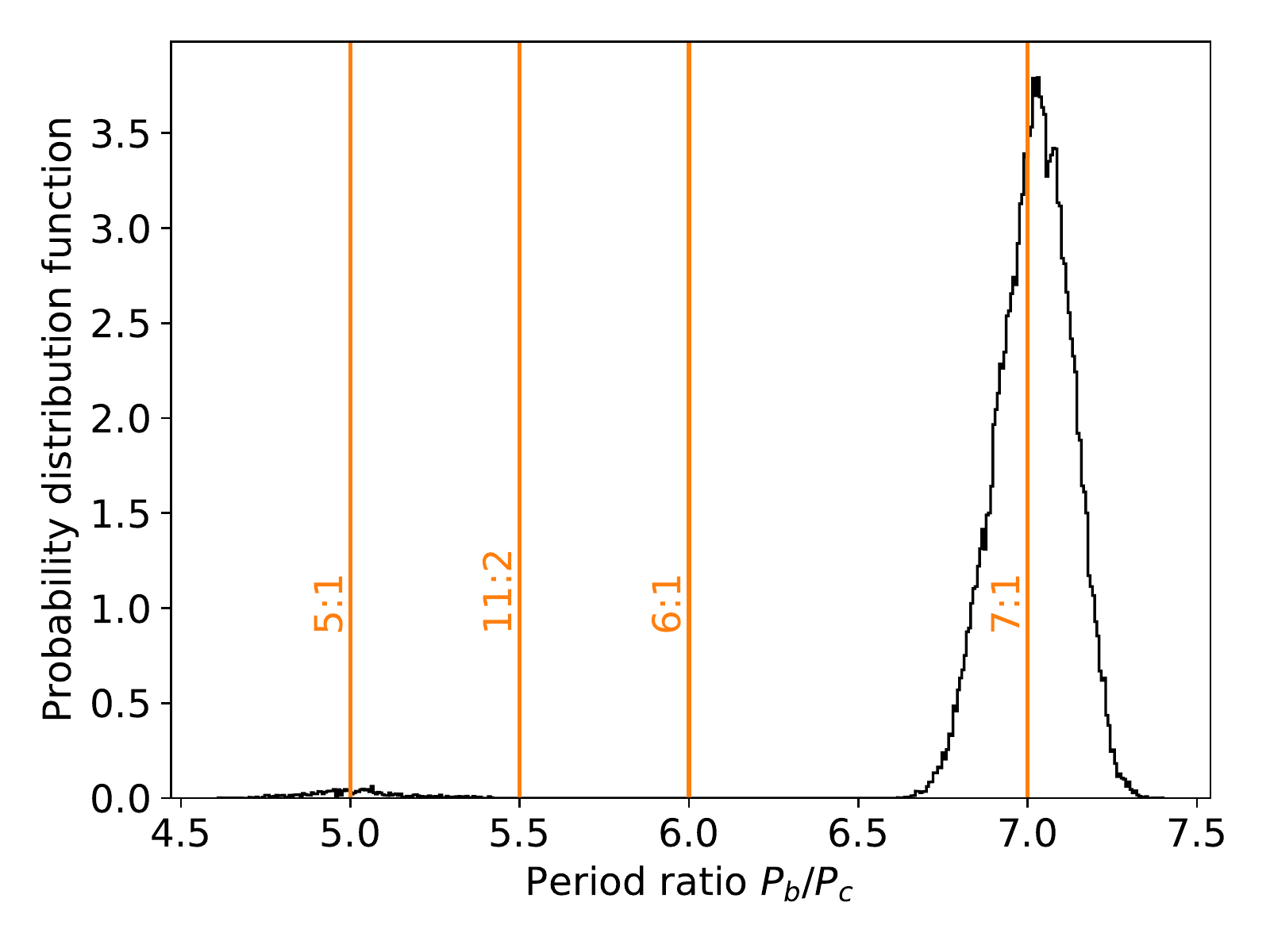}
    \caption{Period ratio $P_b/P_c$ distribution from the fit using all data (Sect. \ref{sec:full}). If the system is in MMR, the most likely configuration would be the $7:1$ commensurability.}
    \label{fig:MMR}
\end{figure}

Both HR 8799 and PDS 70 planets are suspected to be in a configuration of mean-motion resonance \citep[MMR; respectively $1:2:4:8$ and $2:1$;][]{wangDynamicalConstraintsHR2018,wangConstrainingNaturePDS2021}. Although the orbital fits are not yet precise enough to confirm this, commensurable periods are compatible with the astrometric constraints and increase the stability of the systems. In the $\beta$ Pictoris system, however, the separation between the planets forbids any low-order MMR. Figure~\ref{fig:MMR} points to a possible $7:1$ configuration. Such a high-order MMR is not expected to have an impact on the dynamics of the planets, especially if we consider the high masses of the planets that will introduce non-negligible short-scale variations to the orbital elements.

The existence and characteristics of $\beta$ Pictoris b had notably been postulated long before the first direct-imaging detection as a consequence of the observation of numerous falling evaporating bodies (FEBs)
close to the star \citep{beustMeanMotionResonancesSource1996,beustFallingEvaporatingBodies2000}. Those star-grazing
planetesimals were thought to come from an asteroid belt in 4 : 1
MMR with a slightly eccentric giant planet with predicted orbital characteristics that approximately match those of  $\beta$ Pictoris b \citep{2001A&A...376..621T,2007A&A...466..201B}.
The presence of another massive planet such as  $\beta$ Pictoris c orbiting well inside $\beta$ Pictoris b can perturb this picture. As a matter of fact, FEB progenitors trapped in 4:1 MMR with $\beta$ Pictoris b (i.e., orbiting at $\sim$4 au from the star) undergo a gradual eccentricity increase with only little changes in semimajor axis before reaching the star-grazing state. This way they inevitably come to cross the orbit of  $\beta$ Pictoris c twice per orbit. Thus, close encounters with that planet would presumably eject most of the comets before they can become observable FEBs. Hopefully, the refined orbital and mass constraints provided in the present paper will help explore this rich dynamical issue. Beust et al. (2021, in preparation) show that the presence of $\beta$ Pictoris c entirely prevents FEB progenitors from reaching the FEB state. Depending on its eccentricity, their source 4:1 MMR at 4 au may not even be stable. However, various inner MMRs with $\beta$ Pictoris c (instead of b) could be a valuable alternate source of FEBs, leading to a revision of the FEB scenario.

\section{Conclusions}
\label{sec:conc}

Our results demonstrate the capabilities of relative astrometry with optical interferometry to characterize multi-planetary systems:
   \begin{enumerate}
    \item  We were able to detect  $\beta$ Pictoris c  and measure its mass from the astrometry of $\beta$ Pictoris b alone. Our \textrm{mass measurement of $10.04^{+4.53}_{-3.10}\,M_{\rm Jup}$ is} in agreement with the previously published estimations based on RV by
      \citet{lagrangeUnveilingPictorisSystem2020}
      and \citet{vandalDynamicalMassEstimates2020}. This is the first time that the presence of an exoplanet has been derived from the astrometry of another exoplanet.
      \item The new detection of $\beta$ Pictoris c, at 96\,mas from its star, is a record in terms of exoplanet detection at small separation. Thanks to this new measurement, the orbital parameters of c, especially its eccentricity at $0.32\pm0.02$, are well determined.
       Fitting all data sets together, we obtain $8.89 \pm 0.75\,M_{\rm Jup}$ for $\beta$ Pictoris c and
       $11.90^{+2.93}_{-3.04}\,M_{\rm Jup}$ for $\beta$ Pictoris b.
      \item Due to their masses and eccentricities, $\beta$ Pictoris b and c are strongly interacting on secular timescales. The eccentricity of $\beta$ Pictoris b may not be primordial but may simply be due to secular perturbations arising from $\beta$ Pictoris c. The orbital parameters of the two planets point toward a possible 7 : 1 MMR.
   \end{enumerate}

\begin{acknowledgements}
Letter based on observations collected at the European Organisation for Astronomical Research in the Southern Hemisphere, ID 1104.C-0651. This research has made use of the Jean-Marie Mariotti Center \texttt{Aspro} service  (\url{http://www.jmmc.fr/aspro}) .
This research also made use of Astropy (\url{http://www.astropy.org}), a community-developed core Python package for Astronomy \citep{2018AJ....156..123T}.
 J.J.W. is supported by the Heising-Simons Foundation 51 Pegasi b Fellowship. The development of \texttt{orbitize!} is supported by the Heising-Simons Foundation through grant 2019-1698. We acknowledge support from the European Research Council under the Horizon 2020 Framework Program via the ERC grants 832428 (T.H.) and 757561 (A.V. and G.P.P.O.). A.A. and P.G. were supported by Funda\c{c}\~{a}o para a Ci\^{e}ncia e a Tecnologia, with grants reference UIDB/00099/2020 and  PTDC/FIS-AST/7002/2020.
\end{acknowledgements}

\begin{appendix}


\section{Log of observations}

The log of the three observations of $\beta$ Pictoris system is presented in Table~\ref{tab:log}.

\begin{table*}
  \begin{center}
    \begin{tabular}{cccccccc}
      \hline
      \hline
      Date & Target & UT Start Time &UT End Time & $N$exp / $N$DIT / DIT & airmass & $\tau_0$ & seeing \\
      \hline
      6 January 2021 & $\beta$ Pictoris c & 02:32:52 &04:01:17 & 18 / 32 / 10\,s&  1.12-1.18   &  7-13\,ms & 0.4-0.5"  \\
       7 January 2021 & $\beta$ Pictoris b & 05:17:34 & 06:00:41 & 7  / 32 / 10\,s &  1.21-1.32  & 5-11\,ms & 0.4-0.7" \\
       27 August 2021 & $\beta$ Pictoris b & 09:00:31 & 09:46:39 & 5  / 16 / 30\,s &  1.24-1.39  & 2-3\,ms & 1.0-1.7" \\
      \hline
      \hline
    \end{tabular}
    \caption{Log of the two new GRAVITY observations. $N$exp is the number of exposures. $N$DIT and DIT are the number and duration of integrations per exposure. $\tau_0$ is the coherence time.}
    \label{tab:log}
  \end{center}
\end{table*}

\section{N-body code versus Jacobi approximation}\label{sec:nbody}

\begin{figure*}
      \resizebox{\hsize}{!}{\includegraphics{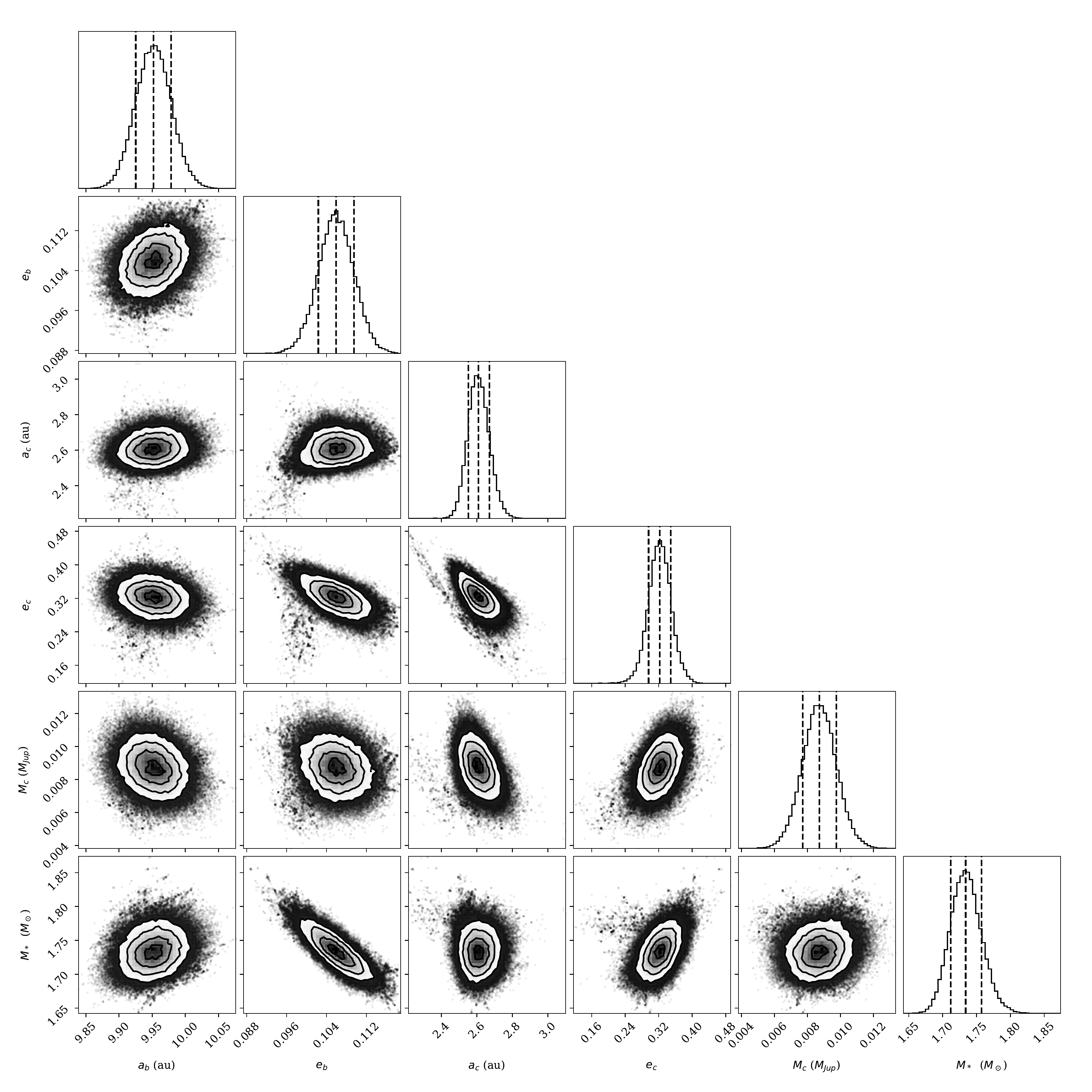}}
      \caption{Corner plot of selected orbital parameters of interest for  the two-planet model MCMC fit of the $\beta$ Pictoris b astrometry only.}
     \label{fig:corner}
    \end{figure*}

A corner plot of the key parameters obtained from the two-planet fit with b and c astrometry is shown in Fig.~\ref{fig:corner}.

To assess the validity of our orbit model for this two-planet fit with b and c astrometry, we used it to fit simulated astrometry of the $\beta$ Pictoris system generated by an $N$-body integrator where we know the truth values. We used the \texttt{REBOUND} package with its IAS15 integrator to model the three-body system \citep{Rein2012,Rein2015}. We initialized the two planets at MJD 59,000 using the osculating orbital elements defined in the Truth column of Table \ref{tab:simResults}. We then integrated the planets forward and backward in time to simulate astrometry at every epoch where we have real astrometric data. For each epoch, random Gaussian noise was added to each measurement based on the corresponding reported measurement uncertainty at that epoch. We then estimated the orbital parameters with \texttt{orbitize!}, using the same procedure as above when fitting all of the astrometric data on both planets.

\begin{figure*}
  \resizebox{\hsize}{!}{\includegraphics{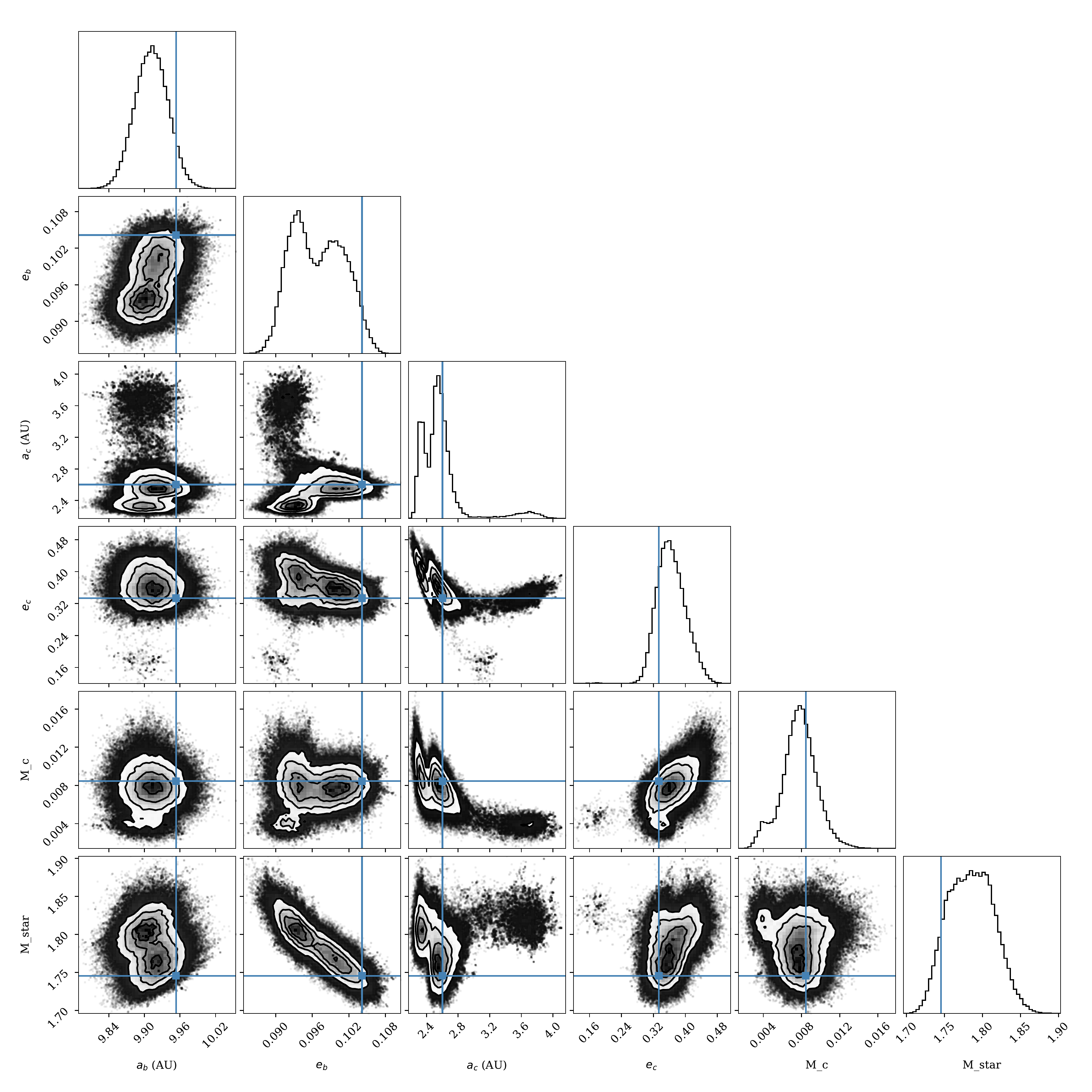}}
  \caption{Corner plot of selected parameters of interest from the fit to simulated astrometry. Truth values of each parameter are plotted as cyan lines. }
  \label{fig:rebound_corner}
\end{figure*}

The median and 68\% and 95\% credible intervals are listed in Table \ref{tab:simResults}. The posteriors, as well as the truth values, are also plotted in Fig. \ref{fig:rebound_corner}. We find that the orbital parameters and mass of $\beta$ Pictoris c derived from the \texttt{orbitize!} model is accurate to within the 68\% credible interval. The orbital elements of $\beta$ Pictoris b sometimes appear beyond the 68\% credible interval, indicating that the orbital model may have some biases in estimating the osculating orbital elements of planet b. Overall, we find the orbital model to be sufficiently accurate, especially for deriving a dynamical mass of $\beta$ Pictoris c.

\begin{table}
    \centering
    \begin{tabular}{c|c|c}
        \hline
        \hline
        Parameter & Truth & Fit Credible Interval \\
        \hline
        $a_b$ & 9.95 & $9.90^{+0.03(+0.06)}_{-0.03(-0.06)}$ \\
        $e_b$ & 0.10 & $0.09^{+0.01(+0.01)}_{+0.00(+0.00)}$ \\
        $i_b$ & 88.98 & $89.00^{+0.01(+0.03)}_{-0.01(-0.02)}$ \\
        $\omega_b$ & 202.05 & $189.25^{+7.07(+10.96)}_{-7.15(-15.58)}$ \\
        $\Omega_b$ & 31.81 & $31.81^{+0.01(+0.02)}_{-0.01(-0.02)}$ \\
        $\tau_b$ & 0.73 & $0.69^{+0.02(+0.03)}_{-0.03(-0.05)}$ \\
        \hline
        $a_c$ & 2.60 & $2.55^{+0.90(+1.19)}_{-0.19(-0.27)}$ \\
        $e_c$ & 0.33 & $0.37^{+0.05(+0.09)}_{-0.03(-0.06)}$ \\
        $i_c$ & 88.82 & $88.83^{+0.19(+0.34)}_{-0.19(-0.38)}$ \\
        $\omega_c$ & 61.02 & $68.40^{+253.47(+257.53)}_{-11.18(-14.63)}$ \\
        $\Omega_c$ & 31.06 & $31.05^{+0.84(+1.06)}_{-0.07(-0.14)}$ \\
        $\tau_c$ & 0.71 & $0.71^{+0.14(+0.17)}_{-0.31(-0.32)}$ \\
        \hline
        Parallax & 51.39 & $51.44^{+0.12(+0.24)}_{-0.12(-0.24)}$ \\
        \hline
        $M_b$ & 10.00 & $10.00$ \\
        $M_c$ & 8.85 & $7.69^{+2.04(+4.32)}_{-3.42(-4.66)}$ \\
        $M_*$ & 1.75 & $1.81^{+0.05(+0.09)}_{-0.04(-0.07)}$ \\
      \hline
      \hline
    \end{tabular}
    \caption{Orbital parameters derived from fitting to simulated astrometry. The units are the same as in Table \ref{tab:allResultsl}.}
    \label{tab:simResults}
\end{table}

\section{Secular evolution}\label{sec:secular}

\begin{figure}[!h]
    \centering
    \includegraphics[width=\linewidth]{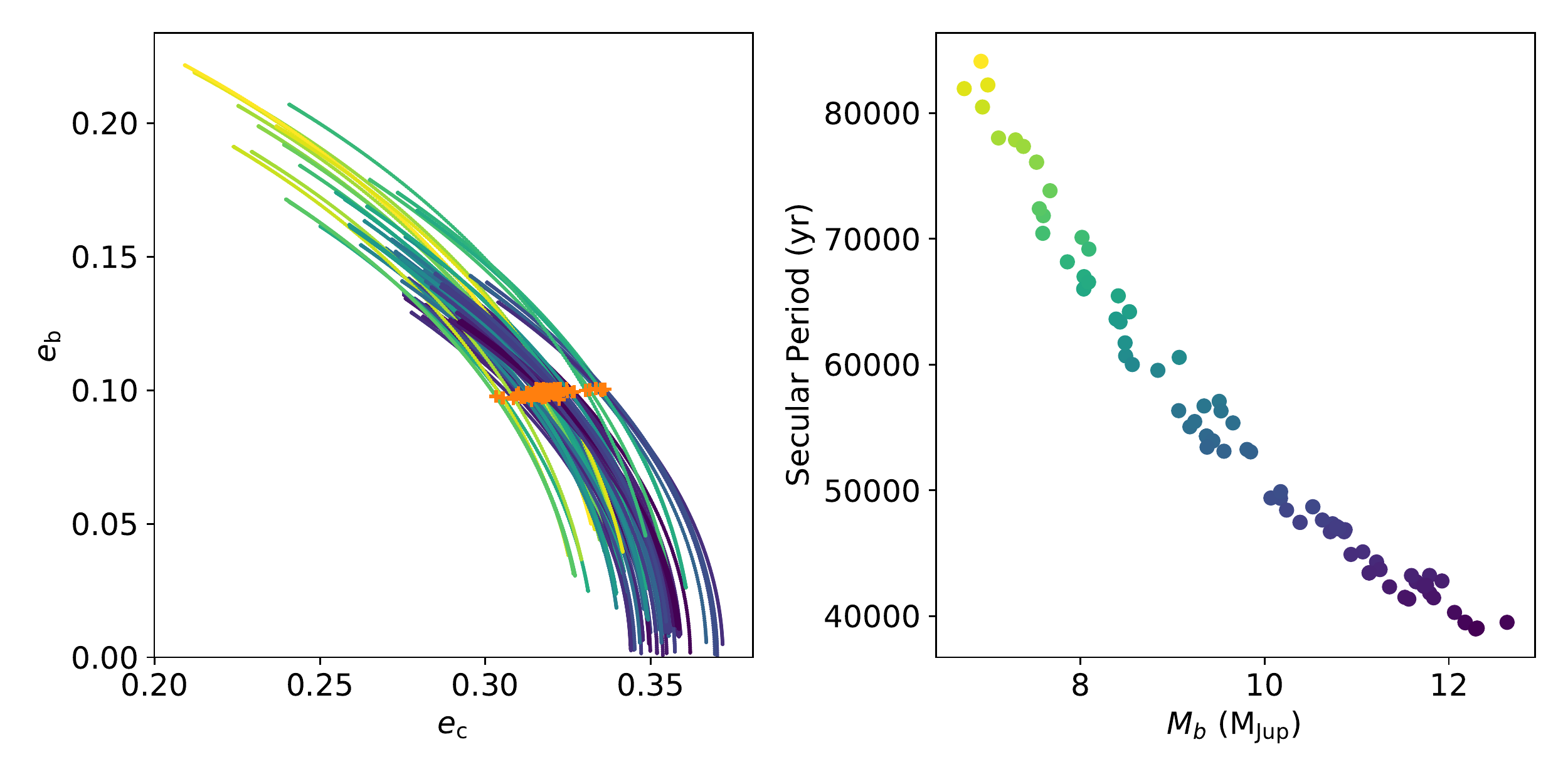}
    \caption{Secular evolution of the eccentricity of b and c due to their mutual interaction. Left is the trajectory in eccentricity phase space, starting from the observed values (orange crosses, which are random solutions to the orbital fit using all available data). The colors represent the corresponding eccentricity evolution period. The right panel represents the aforementioned period with respect to the mass of b.}
    \label{fig:sec}
\end{figure}

Two planets orbiting a star mutually interact with each other. If they are in a stable configuration, their orbital elements will undergo variations on a timescale significantly larger than the orbital periods, the so-called secular variations. In a coplanar system (which we can consider $\beta$ Pictoris to be), the evolution of the eccentricity can be derived analytically \citep[Eq.~7.28 of][]{murraySolarSystemDynamics2000}. For each solution of the orbital fits, we can then compute the period of the eccentricity oscillations and the range of eccentricities that the planets will reach within this period (see Fig.~\ref{fig:sec}).

\end{appendix}

\bibliographystyle{aa}
\bibliography{MyLibrary,JasonLib,LaetitiaLib,astropy,Beust,GillesLib}

\begin{thebibliography}{62}
\expandafter\ifx\csname natexlab\endcsname\relax\def\natexlab#1{#1}\fi

\bibitem[{Agol {et~al.}(2005)Agol, Steffen, Sari, \&
  Clarkson}]{agolDetectingTerrestrialPlanets2005}
Agol, E., Steffen, J., Sari, R., \& Clarkson, W. 2005, Monthly Notices of the
  Royal Astronomical Society, 359, 567

\bibitem[{Alibert {et~al.}(2005)Alibert, Mordasini, Benz, \&
  Winisdoerffer}]{alibertModelsGiantPlanet2005}
Alibert, Y., Mordasini, C., Benz, W., \& Winisdoerffer, C. 2005, Astronomy and
  Astrophysics, 434, 343

\bibitem[{Arago(1846)}]{Arago:1846}
Arago, F. 1846, in Comptes rendus hebdomadaires des s\'eances de l'Acad\'emie
  des sciences, Vol.~23 (Acad\'emie des sciences (France)), 659--662,
  \url{https://gallica.bnf.fr/ark:/12148/bpt6k2980r/f663}

\bibitem[{{Augereau} {et~al.}(2001){Augereau}, {Nelson}, {Lagrange},
  {Papaloizou}, \& {Mouillet}}]{2001A&A...370..447A}
{Augereau}, J.~C., {Nelson}, R.~P., {Lagrange}, A.~M., {Papaloizou}, J.~C.~B.,
  \& {Mouillet}, D. 2001, \aap, 370, 447

\bibitem[{Beust(2003)}]{beustSymplecticIntegrationHierarchical2003}
Beust, H. 2003, Astronomy and Astrophysics, 400, 1129

\bibitem[{Beust \& Morbidelli(1996)}]{beustMeanMotionResonancesSource1996}
Beust, H. \& Morbidelli, A. 1996, Icarus, 120, 358

\bibitem[{Beust \& Morbidelli(2000)}]{beustFallingEvaporatingBodies2000}
Beust, H. \& Morbidelli, A. 2000, \i carus, 143, 170

\bibitem[{{Beust} \& {Valiron}(2007)}]{2007A&A...466..201B}
{Beust}, H. \& {Valiron}, P. 2007, \aap, 466, 201

\bibitem[{{Blunt} {et~al.}(2020){Blunt}, {Wang}, {Angelo}, {Ngo}, {Cody}, {De
  Rosa}, {Graham}, {Hirsch}, {Nagpal}, {Nielsen}, {Pearce}, {Rice}, \&
  {Tejada}}]{Blunt2020}
{Blunt}, S., {Wang}, J.~J., {Angelo}, I., {et~al.} 2020, \aj, 159, 89

\bibitem[{{Bohn} {et~al.}(2020){Bohn}, {Kenworthy}, {Ginski}, {Rieder},
  {Mamajek}, {Meshkat}, {Pecaut}, {Reggiani}, {de Boer}, {Keller}, {Snik}, \&
  {Southworth}}]{2020ApJ...898L..16B}
{Bohn}, A.~J., {Kenworthy}, M.~A., {Ginski}, C., {et~al.} 2020, \apjl, 898, L16

\bibitem[{Boss(2011)}]{bossFormationGiantPlanets2011}
Boss, A.~P. 2011, The Astrophysical Journal, 731, 74

\bibitem[{Bowler {et~al.}(2020)Bowler, Blunt, \&
  Nielsen}]{bowlerPopulationlevelEccentricityDistributions2020}
Bowler, B.~P., Blunt, S.~C., \& Nielsen, E.~L. 2020, The Astronomical Journal,
  159, 63

\bibitem[{{Brandt} {et~al.}(2021{\natexlab{a}}){Brandt}, {Brandt}, {Dupuy},
  {Li}, \& {Michalik}}]{Brandt2021betapic}
{Brandt}, G.~M., {Brandt}, T.~D., {Dupuy}, T.~J., {Li}, Y., \& {Michalik}, D.
  2021{\natexlab{a}}, \aj, 161, 179

\bibitem[{{Brandt} {et~al.}(2021{\natexlab{b}}){Brandt}, {Brandt}, {Dupuy},
  {Michalik}, \& {Marleau}}]{2021ApJ...915L..16B}
{Brandt}, G.~M., {Brandt}, T.~D., {Dupuy}, T.~J., {Michalik}, D., \& {Marleau},
  G.-D. 2021{\natexlab{b}}, \apjl, 915, L16

\bibitem[{{Bryan} {et~al.}(2019){Bryan}, {Knutson}, {Lee}, {Fulton}, {Batygin},
  {Ngo}, \& {Meshkat}}]{Bryan2019}
{Bryan}, M.~L., {Knutson}, H.~A., {Lee}, E.~J., {et~al.} 2019, \aj, 157, 52

\bibitem[{{Christiansen} {et~al.}(2017){Christiansen}, {Vanderburg}, {Burt},
  {Fulton}, {Batygin}, {Benneke}, {Brewer}, {Charbonneau}, {Ciardi}, {Collier
  Cameron}, {Coughlin}, {Crossfield}, {Dressing}, {Greene}, {Howard}, {Latham},
  {Molinari}, {Mortier}, {Mullally}, {Pepe}, {Rice}, {Sinukoff}, {Sozzetti},
  {Thompson}, {Udry}, {Vogt}, {Barman}, {Batalha}, {Bouchy}, {Buchhave},
  {Butler}, {Cosentino}, {Dupuy}, {Ehrenreich}, {Fiorenzano}, {Hansen},
  {Henning}, {Hirsch}, {Holden}, {Isaacson}, {Johnson}, {Knutson}, {Kosiarek},
  {L{\'o}pez-Morales}, {Lovis}, {Malavolta}, {Mayor}, {Micela}, {Motalebi},
  {Petigura}, {Phillips}, {Piotto}, {Rogers}, {Sasselov}, {Schlieder},
  {S{\'e}gransan}, {Watson}, \& {Weiss}}]{Christiansen2017}
{Christiansen}, J.~L., {Vanderburg}, A., {Burt}, J., {et~al.} 2017, \aj, 154,
  122

\bibitem[{{Demory} {et~al.}(2020){Demory}, {Pozuelos}, {G{\'o}mez Maqueo Chew},
  {Sabin}, {Petrucci}, {Schroffenegger}, {Grimm}, {Sestovic}, {Gillon},
  {McCormac}, {Barkaoui}, {Benz}, {Bieryla}, {Bouchy}, {Burdanov}, {Collins},
  {de Wit}, {Dressing}, {Garcia}, {Giacalone}, {Guerra}, {Haldemann}, {Heng},
  {Jehin}, {Jofr{\'e}}, {Kane}, {Lillo-Box}, {Maign{\'e}}, {Mordasini},
  {Morris}, {Niraula}, {Queloz}, {Rackham}, {Savel}, {Soubkiou}, {Srdoc},
  {Stassun}, {Triaud}, {Zambelli}, {Ricker}, {Latham}, {Seager}, {Winn},
  {Jenkins}, {Calvario-Vel{\'a}squez}, {Franco Herrera}, {Colorado}, {Cadena
  Zepeda}, {Figueroa}, {Watson}, {Lugo-Ibarra}, {Carigi}, {Guisa}, {Herrera},
  {Sierra D{\'\i}az}, {Su{\'a}rez}, {Barrado}, {Batalha}, {Benkhaldoun},
  {Chontos}, {Dai}, {Essack}, {Ghachoui}, {Huang}, {Huber}, {Isaacson},
  {Lissauer}, {Morales-Calder{\'o}n}, {Robertson}, {Roy}, {Twicken},
  {Vanderburg}, \& {Weiss}}]{2020A&A...642A..49D}
{Demory}, B.~O., {Pozuelos}, F.~J., {G{\'o}mez Maqueo Chew}, Y., {et~al.} 2020,
  \aap, 642, A49

\bibitem[{{Dupuy} {et~al.}(2019){Dupuy}, {Brandt}, {Kratter}, \&
  {Bowler}}]{2019ApJ...871L...4D}
{Dupuy}, T.~J., {Brandt}, T.~D., {Kratter}, K.~M., \& {Bowler}, B.~P. 2019,
  \apjl, 871, L4

\bibitem[{Emsenhuber {et~al.}(2020)Emsenhuber, Mordasini, Burn, Alibert, Benz,
  \& Asphaug}]{emsenhuberNewGenerationPlanetary2020a}
Emsenhuber, A., Mordasini, C., Burn, R., {et~al.} 2020, arXiv e-prints, 2007,
  arXiv:2007.05562

\bibitem[{{Foreman-Mackey} {et~al.}(2013){Foreman-Mackey}, Hogg, Lang, \&
  Goodman}]{foreman-mackeyEmceeMCMCHammer2013}
{Foreman-Mackey}, D., Hogg, D.~W., Lang, D., \& Goodman, J. 2013, Publications
  of the Astronomical Society of the Pacific, 125, 306

\bibitem[{{Gravity Collaboration} {et~al.}(2017){Gravity Collaboration},
  Abuter, Accardo, Amorim, Anugu, {\'A}vila, Azouaoui, Benisty, Berger, Blind,
  Bonnet, Bourget, Brandner, Brast, Buron, Burtscher, Cassaing, Chapron,
  Choquet, Cl{\'e}net, Collin, Coud{\'e} Du~Foresto, {de Wit}, {de Zeeuw},
  Deen, {Delplancke-Str{\"o}bele}, Dembet, Derie, Dexter, Duvert, Ebert,
  Eckart, Eisenhauer, Esselborn, F{\'e}dou, Finger, Garcia, Garcia~Dabo,
  Garcia~Lopez, Gendron, Genzel, Gillessen, Gonte, Gordo, Grould,
  Gr{\"o}zinger, Guieu, Haguenauer, Hans, Haubois, Haug, Haussmann, Henning,
  Hippler, Horrobin, Huber, Hubert, Hubin, Hummel, Jakob, Janssen, Jochum,
  Jocou, Kaufer, Kellner, Kendrew, Kern, Kervella, Kiekebusch, Klein, Kok,
  Kolb, Kulas, Lacour, Lapeyr{\`e}re, Lazareff, Le~Bouquin, L{\`e}na, Lenzen,
  L{\'e}v{\^e}que, Lippa, Magnard, Mehrgan, Mellein, M{\'e}rand,
  {Moreno-Ventas}, Moulin, M{\"u}ller, M{\"u}ller, Neumann, Oberti, Ott,
  Pallanca, Panduro, Pasquini, Paumard, Percheron, Perraut, Perrin,
  Pfl{\"u}ger, Pfuhl, Phan~Duc, Plewa, Popovic, Rabien, Ram{\'i}rez, Ramos,
  Rau, Riquelme, Rohloff, Rousset, {Sanchez-Bermudez}, Scheithauer,
  Sch{\"o}ller, Schuhler, Spyromilio, Straubmeier, Sturm, Suarez, Tristram,
  Ventura, Vincent, Waisberg, Wank, Weber, Wieprecht, Wiest, Wiezorrek,
  Wittkowski, Woillez, Wolff, Yazici, Ziegler, \&
  Zins}]{gravitycollaborationFirstLightGRAVITY2017}
{Gravity Collaboration}, Abuter, R., Accardo, M., {et~al.} 2017, Astronomy and
  Astrophysics, 602, A94

\bibitem[{{Gravity Collaboration} {et~al.}(2018){Gravity Collaboration},
  Abuter, Amorim, Baub{\"o}ck, Berger, Bonnet, Brandner, Cl{\'e}net, Coud{\'e}
  Du~Foresto, {de Zeeuw}, Deen, Dexter, Duvert, Eckart, Eisenhauer,
  F{\"o}rster~Schreiber, Garcia, Gao, Gendron, Genzel, Gillessen, Guajardo,
  Habibi, Haubois, Henning, Hippler, Horrobin, Huber, {Jim{\'e}nez-Rosales},
  Jocou, Kervella, Lacour, Lapeyr{\`e}re, Lazareff, Le~Bouquin, L{\'e}na,
  Lippa, Ott, Panduro, Paumard, Perraut, Perrin, Pfuhl, Plewa, Rabien,
  {Rodr{\'i}guez-Coira}, Rousset, Sternberg, Straub, Straubmeier, Sturm,
  Tacconi, Vincent, {von Fellenberg}, Waisberg, Widmann, Wieprecht, Wiezorrek,
  Woillez, \& Yazici}]{gravitycollaborationDetectionOrbitalMotions2018}
{Gravity Collaboration}, Abuter, R., Amorim, A., {et~al.} 2018, Astronomy and
  Astrophysics, 618, L10

\bibitem[{{GRAVITY Collaboration} {et~al.}(2021){GRAVITY Collaboration},
  Abuter, Amorim, Baub{\"o}ck, Berger, Bonnet, Brandner, Cl{\'e}net, Davies,
  {de Zeeuw}, Dexter, Dallilar, Drescher, Eckart, Eisenhauer,
  F{\"o}rster~Schreiber, Garcia, Gao, Gendron, Genzel, Gillessen, Habibi,
  Haubois, Hei{\ss}el, Henning, Hippler, Horrobin, {Jim{\'e}nez-Rosales},
  Jochum, Jocou, Kaufer, Kervella, Lacour, Lapeyr{\`e}re, Le~Bouquin, L{\'e}na,
  Lutz, Nowak, Ott, Paumard, Perraut, Perrin, Pfuhl, Rabien,
  {Rodr{\'i}guez-Coira}, Shangguan, Shimizu, Scheithauer, Stadler, Straub,
  Straubmeier, Sturm, Tacconi, Vincent, {von Fellenberg}, Waisberg, Widmann,
  Wieprecht, Wiezorrek, Woillez, Yazici, Young, \&
  Zins}]{gravitycollaborationImprovedGRAVITYAstrometric2021}
{GRAVITY Collaboration}, Abuter, R., Amorim, A., {et~al.} 2021, A\&A, 647, A59

\bibitem[{{Gravity Collaboration} {et~al.}(2019){Gravity Collaboration},
  Lacour, Nowak, Wang, Pfuhl, Eisenhauer, Abuter, Amorim, Anugu, Benisty,
  Berger, Beust, Blind, Bonnefoy, Bonnet, Bourget, Brandner, Buron, Collin,
  Charnay, Chapron, Cl{\'e}net, Coud{\'e} Du~Foresto, {de Zeeuw}, Deen, Dembet,
  Dexter, Duvert, Eckart, F{\"o}rster~Schreiber, F{\'e}dou, Garcia,
  Garcia~Lopez, Gao, Gendron, Genzel, Gillessen, Gordo, Greenbaum, Habibi,
  Haubois, Hau{\ss}mann, Henning, Hippler, Horrobin, Hubert, Jimenez~Rosales,
  Jocou, Kendrew, Kervella, Kolb, Lagrange, Lapeyr{\`e}re, Le~Bouquin,
  L{\'e}na, Lippa, Lenzen, Maire, Molli{\`e}re, Ott, Paumard, Perraut, Perrin,
  Pueyo, Rabien, Ram{\'i}rez, Rau, {Rodr{\'i}guez-Coira}, Rousset,
  {Sanchez-Bermudez}, Scheithauer, Schuhler, Straub, Straubmeier, Sturm,
  Tacconi, Vincent, {van Dishoeck}, {von Fellenberg}, Wank, Waisberg, Widmann,
  Wieprecht, Wiest, Wiezorrek, Woillez, Yazici, Ziegler, \&
  Zins}]{gravitycollaborationFirstDirectDetection2019}
{Gravity Collaboration}, Lacour, S., Nowak, M., {et~al.} 2019, Astronomy and
  Astrophysics, 623, L11

\bibitem[{{Gravity Collaboration} {et~al.}(2020){Gravity Collaboration}, Nowak,
  Lacour, Molli{\`e}re, Wang, Charnay, {van Dishoeck}, Abuter, Amorim, Berger,
  Beust, Bonnefoy, Bonnet, Brandner, Buron, Cantalloube, Collin, Chapron,
  Cl{\'e}net, Coud{\'e} Du~Foresto, {de Zeeuw}, Dembet, Dexter, Duvert, Eckart,
  Eisenhauer, F{\"o}rster~Schreiber, F{\'e}dou, Garcia~Lopez, Gao, Gendron,
  Genzel, Gillessen, Hau{\ss}mann, Henning, Hippler, Hubert, Jocou, Kervella,
  Lagrange, Lapeyr{\`e}re, Le~Bouquin, L{\'e}na, Maire, Ott, Paumard, Paladini,
  Perraut, Perrin, Pueyo, Pfuhl, Rabien, Rau, {Rodr{\'i}guez-Coira}, Rousset,
  Scheithauer, Shangguan, Straub, Straubmeier, Sturm, Tacconi, Vincent,
  Widmann, Wieprecht, Wiezorrek, Woillez, Yazici, \&
  Ziegler}]{gravitycollaborationPeeringFormationHistory2020}
{Gravity Collaboration}, Nowak, M., Lacour, S., {et~al.} 2020, Astronomy and
  Astrophysics, 633, A110

\bibitem[{Haffert {et~al.}(2019)Haffert, Bohn, {de Boer}, Snellen, Brinchmann,
  Girard, Keller, \& Bacon}]{haffertTwoAccretingProtoplanets2019}
Haffert, S.~Y., Bohn, A.~J., {de Boer}, J., {et~al.} 2019, Nature Astronomy

\bibitem[{{Heap} {et~al.}(2000){Heap}, {Lindler}, {Lanz}, {Cornett}, {Hubeny},
  {Maran}, \& {Woodgate}}]{2000ApJ...539..435H}
{Heap}, S.~R., {Lindler}, D.~J., {Lanz}, T.~M., {et~al.} 2000, \apj, 539, 435

\bibitem[{Keppler {et~al.}(2018)Keppler, Benisty, M{\"u}ller, Henning, {van
  Boekel}, Cantalloube, Ginski, {van Holstein}, Maire, Pohl, Samland, Avenhaus,
  Baudino, Boccaletti, {de Boer}, Bonnefoy, Chauvin, Desidera, Langlois,
  Lazzoni, Marleau, Mordasini, Pawellek, Stolker, Vigan, Zurlo, Birnstiel,
  Brandner, Feldt, Flock, Girard, Gratton, Hagelberg, Isella, Janson, Juhasz,
  Kemmer, Kral, Lagrange, Launhardt, Matter, M{\'e}nard, Milli, Molli{\`e}re,
  Olofsson, P{\'e}rez, Pinilla, Pinte, Quanz, Schmidt, Udry, Wahhaj, Williams,
  Buenzli, Cudel, Dominik, Galicher, Kasper, Lannier, Mesa, Mouillet, Peretti,
  Perrot, Salter, Sissa, Wildi, Abe, Antichi, Augereau, Baruffolo, Baudoz,
  Bazzon, Beuzit, Blanchard, Brems, Buey, De~Caprio, Carbillet, Carle, Cascone,
  Cheetham, Claudi, Costille, Delboulb{\'e}, Dohlen, Fantinel, Feautrier,
  Fusco, Giro, Gluck, Gry, Hubin, Hugot, Jaquet, Le~Mignant, Llored, Madec,
  Magnard, Martinez, Maurel, Meyer, {M{\"o}ller-Nilsson}, Moulin, Mugnier,
  Orign{\'e}, Pavlov, Perret, Petit, Pragt, Puget, Rabou, Ramos, Rigal, Rochat,
  Roelfsema, Rousset, Roux, Salasnich, Sauvage, Sevin, Soenke, Stadler, Suarez,
  Turatto, \& Weber}]{kepplerDiscoveryPlanetarymassCompanion2018}
Keppler, M., Benisty, M., M{\"u}ller, A., {et~al.} 2018, Astronomy and
  Astrophysics, 617, A44

\bibitem[{{Konacki} \& {Wolszczan}(2003)}]{2003ApJ...591L.147K}
{Konacki}, M. \& {Wolszczan}, A. 2003, \apjl, 591, L147

\bibitem[{Lacour {et~al.}(2019)Lacour, Dembet, Abuter, F{\'e}dou, Perrin,
  Choquet, Pfuhl, Eisenhauer, Woillez, Cassaing, Wieprecht, Ott, Wiezorrek,
  Tristram, Wolff, Ram{\'i}rez, Haubois, Perraut, Straubmeier, Brandner, \&
  Amorim}]{lacourGRAVITYFringeTracker2019}
Lacour, S., Dembet, R., Abuter, R., {et~al.} 2019, A\&A, 624, A99

\bibitem[{Lacour {et~al.}(2014)Lacour, Eisenhauer, Gillessen, Pfuhl, Woillez,
  Bonnet, Perrin, Lazareff, Rabien, Lapeyr{\`e}re, Cl{\'e}net, Kervella, \&
  Kok}]{lacourReachingMicroarcsecondAstrometry2014}
Lacour, S., Eisenhauer, F., Gillessen, S., {et~al.} 2014, Astronomy and
  Astrophysics, 567, A75

\bibitem[{{Lacour} {et~al.}(2020){Lacour}, {Wang}, {Nowak}, {Pueyo},
  {Eisenhauer}, {Lagrange}, {Molli{\`e}re}, {Abuter}, {Amorin},
  {Asensio-Torres}, {Baub{\"o}ck}, {Benisty}, {Berger}, {Beust}, {Blunt},
  {Boccaletti}, {Bohn}, {Bonnefoy}, {Bonnet}, {Brandner}, {Cantalloube},
  {Caselli}, {Charnay}, {Chauvin}, {Choquet}, {Christiaens}, {Cl{\'e}net},
  {Cridland}, {de Zeeuw}, {Dembet}, {Dexter}, {Drescher}, {Duvert}, {Gao},
  {Garcia}, {Garcia Lopez}, {Gardner}, {Gendron}, {Genzel}, {Gillessen},
  {Girard}, {Haubois}, {Hei{\ss}el}, {Henning}, {Hinkley}, {Hippler},
  {Horrobin}, {Houll{\'e}}, {Hubert}, {Jim{\'e}nez-Rosales}, {Jocou},
  {Kammerer}, {Keppler}, {Kervella}, {Kreidberg}, {Lapeyr{\`e}re}, {Le
  Bouquin}, {L{\'e}na}, {Lutz}, {Maire}, {M{\'e}rand}, {Monnier}, {Mouillet},
  {Muller}, {Nasedkin}, {Ott}, {Otten}, {Paladini}, {Paumard}, {Perraut},
  {Perrin}, {Pfuhl}, {Rameau}, {Rodet}, {Rodriguez-Coira}, {Rousset},
  {Shangguan}, {Shimizu}, {Stadler}, {Straub}, {Straubmeier}, {Sturm},
  {Stolker}, {van Dishoeck}, {Vigan}, {Vincent}, {von Fellenberg},
  {Ward-Duong}, {Widmann}, {Wieprecht}, {Wiezorrek}, \&
  {Woillez}}]{2020SPIE11446E..0OL}
{Lacour}, S., {Wang}, J.~J., {Nowak}, M., {et~al.} 2020, in Society of
  Photo-Optical Instrumentation Engineers (SPIE) Conference Series, Vol. 11446,
  Society of Photo-Optical Instrumentation Engineers (SPIE) Conference Series,
  114460O

\bibitem[{Lagrange {et~al.}(2010)Lagrange, Bonnefoy, Chauvin, Apai, Ehrenreich,
  Boccaletti, Gratadour, Rouan, Mouillet, Lacour, \&
  Kasper}]{lagrangeGiantPlanetImaged2010a}
Lagrange, A.-M., Bonnefoy, M., Chauvin, G., {et~al.} 2010, Science, 329, 57

\bibitem[{Lagrange {et~al.}(2019)Lagrange, Meunier, Rubini, Keppler, Galland,
  Chapellier, Michel, Balona, Beust, Guillot, Grandjean, Borgniet,
  M{\'e}karnia, Wilson, Kiefer, Bonnefoy, {Lillo-Box}, Pantoja, Jones,
  Iglesias, Rodet, Diaz, Zapata, Abe, \&
  Schmider}]{lagrangeEvidenceAdditionalPlanet2019}
Lagrange, A.-M., Meunier, N., Rubini, P., {et~al.} 2019, Nature Astronomy, 3,
  1135

\bibitem[{{Lagrange} {et~al.}(2020){Lagrange}, {Rubini}, {Nowak}, {Lacour},
  {Grandjean}, {Boccaletti}, {Langlois}, {Delorme}, {Gratton}, {Wang},
  {Flasseur}, {Galicher}, {Kral}, {Meunier}, {Beust}, {Babusiaux}, {Le
  Coroller}, {Thebault}, {Kervella}, {Zurlo}, {Maire}, {Wahhaj}, {Amorim},
  {Asensio-Torres}, {Benisty}, {Berger}, {Bonnefoy}, {Brandner}, {Cantalloube},
  {Charnay}, {Chauvin}, {Choquet}, {Cl{\'e}net}, {Christiaens}, {Coud{\'e} Du
  Foresto}, {de Zeeuw}, {Desidera}, {Duvert}, {Eckart}, {Eisenhauer},
  {Galland}, {Gao}, {Garcia}, {Garcia Lopez}, {Gendron}, {Genzel}, {Gillessen},
  {Girard}, {Hagelberg}, {Haubois}, {Henning}, {Heissel}, {Hippler},
  {Horrobin}, {Janson}, {Kammerer}, {Kenworthy}, {Keppler}, {Kreidberg},
  {Lapeyr{\`e}re}, {Le Bouquin}, {L{\'e}na}, {M{\'e}rand}, {Messina},
  {Molli{\`e}re}, {Monnier}, {Ott}, {Otten}, {Paumard}, {Paladini}, {Perraut},
  {Perrin}, {Pueyo}, {Pfuhl}, {Rodet}, {Rodriguez-Coira}, {Rousset}, {Samland},
  {Shangguan}, {Schmidt}, {Straub}, {Straubmeier}, {Stolker}, {Vigan},
  {Vincent}, {Widmann}, {Woillez}, \& {Gravity
  Collaboration}}]{lagrangeUnveilingPictorisSystem2020}
{Lagrange}, A.~M., {Rubini}, P., {Nowak}, M., {et~al.} 2020, \aap, 642, A18

\bibitem[{{Lapeyrere} {et~al.}(2014){Lapeyrere}, {Kervella}, {Lacour},
  {Azouaoui}, {Garcia-Dabo}, {Perrin}, {Eisenhauer}, {Perraut}, {Straubmeier},
  {Amorim}, \& {Brandner}}]{2014SPIE.9146E..2DL}
{Lapeyrere}, V., {Kervella}, P., {Lacour}, S., {et~al.} 2014, in Society of
  Photo-Optical Instrumentation Engineers (SPIE) Conference Series, Vol. 9146,
  Optical and Infrared Interferometry IV, ed. J.~K. {Rajagopal}, M.~J.
  {Creech-Eakman}, \& F.~{Malbet}, 91462D

\bibitem[{Le~Verrier(1846)}]{LeVerrier:1846}
Le~Verrier, U.-J. 1846, in Comptes rendus hebdomadaires des s\'eances de
  l'Acad\'emie des sciences, Vol.~23 (Acad\'emie des sciences (France)),
  428--438, \url{https://gallica.bnf.fr/ark:/12148/bpt6k2980r/f432}

\bibitem[{Madhusudhan {et~al.}(2017)Madhusudhan, Bitsch, Johansen, \&
  Eriksson}]{madhusudhanAtmosphericSignaturesGiant2017}
Madhusudhan, N., Bitsch, B., Johansen, A., \& Eriksson, L. 2017, Monthly
  Notices of the Royal Astronomical Society, 469, 4102

\bibitem[{Marleau \& Cumming(2014)}]{marleauConstrainingInitialEntropy2014a}
Marleau, G.-D. \& Cumming, A. 2014, Monthly Notices of the Royal Astronomical
  Society, 437, 1378

\bibitem[{Marois {et~al.}(2008)Marois, Macintosh, Barman, Zuckerman, Song,
  Patience, Lafreni{\`e}re, \& Doyon}]{maroisDirectImagingMultiple2008}
Marois, C., Macintosh, B., Barman, T., {et~al.} 2008, Science, 322, 1348

\bibitem[{Marois {et~al.}(2010)Marois, Zuckerman, Konopacky, Macintosh, \&
  Barman}]{maroisImagesFourthPlanet2010}
Marois, C., Zuckerman, B., Konopacky, Q.~M., Macintosh, B., \& Barman, T. 2010,
  Nature, 468, 1080

\bibitem[{{Mordasini}(2013)}]{2013A&A...558A.113M}
{Mordasini}, C. 2013, \aap, 558, A113

\bibitem[{Mordasini {et~al.}(2016)Mordasini, {van Boekel}, Molli{\`e}re,
  Henning, \& Benneke}]{mordasiniImprintExoplanetFormation2016}
Mordasini, C., {van Boekel}, R., Molli{\`e}re, P., Henning, T., \& Benneke, B.
  2016, The Astrophysical Journal, 832, 41

\bibitem[{{Mouillet} {et~al.}(1997){Mouillet}, {Larwood}, {Papaloizou}, \&
  {Lagrange}}]{1997MNRAS.292..896M}
{Mouillet}, D., {Larwood}, J.~D., {Papaloizou}, J.~C.~B., \& {Lagrange}, A.~M.
  1997, \mnras, 292, 896

\bibitem[{M{\"u}ller {et~al.}(2018)M{\"u}ller, Keppler, Henning, Samland,
  Chauvin, Beust, Maire, Molaverdikhani, {van Boekel}, Benisty, Boccaletti,
  Bonnefoy, Cantalloube, Charnay, Baudino, Gennaro, Long, Cheetham, Desidera,
  Feldt, Fusco, Girard, Gratton, Hagelberg, Janson, Lagrange, Langlois,
  Lazzoni, Ligi, M{\'e}nard, Mesa, Meyer, Molli{\`e}re, Mordasini, Moulin,
  Pavlov, Pawellek, Quanz, Ramos, Rouan, Sissa, Stadler, Vigan, Wahhaj, Weber,
  \& Zurlo}]{mullerOrbitalAtmosphericCharacterization2018}
M{\"u}ller, A., Keppler, M., Henning, T., {et~al.} 2018, Astronomy and
  Astrophysics, 617, L2

\bibitem[{Murray \& Dermott(2000)}]{murraySolarSystemDynamics2000}
Murray, C.~D. \& Dermott, S.~F. 2000, Solar {{System Dynamics}} ({Cambridge
  University Press})

\bibitem[{Nayakshin(2017)}]{nayakshinDawesReviewTidal2017}
Nayakshin, S. 2017, Publications of the Astronomical Society of Australia, 34,
  e002

\bibitem[{{Nowak} {et~al.}(2020){Nowak}, {Lacour}, {Lagrange}, {Rubini},
  {Wang}, {Stolker}, {Abuter}, {Amorim}, {Asensio-Torres}, {Baub{\"o}ck},
  {Benisty}, {Berger}, {Beust}, {Blunt}, {Boccaletti}, {Bonnefoy}, {Bonnet},
  {Brandner}, {Cantalloube}, {Charnay}, {Choquet}, {Christiaens}, {Cl{\'e}net},
  {Coud{\'e} Du Foresto}, {Cridland}, {de Zeeuw}, {Dembet}, {Dexter},
  {Drescher}, {Duvert}, {Eckart}, {Eisenhauer}, {Gao}, {Garcia}, {Garcia
  Lopez}, {Gardner}, {Gendron}, {Genzel}, {Gillessen}, {Girard}, {Grandjean},
  {Haubois}, {Hei{\ss}el}, {Henning}, {Hinkley}, {Hippler}, {Horrobin},
  {Houll{\'e}}, {Hubert}, {Jim{\'e}nez-Rosales}, {Jocou}, {Kammerer},
  {Kervella}, {Keppler}, {Kreidberg}, {Kulikauskas}, {Lapeyr{\`e}re}, {Le
  Bouquin}, {L{\'e}na}, {M{\'e}rand}, {Maire}, {Molli{\`e}re}, {Monnier},
  {Mouillet}, {M{\"u}ller}, {Nasedkin}, {Ott}, {Otten}, {Paumard}, {Paladini},
  {Perraut}, {Perrin}, {Pueyo}, {Pfuhl}, {Rameau}, {Rodet},
  {Rodr{\'\i}guez-Coira}, {Rousset}, {Scheithauer}, {Shangguan}, {Stadler},
  {Straub}, {Straubmeier}, {Sturm}, {Tacconi}, {van Dishoeck}, {Vigan},
  {Vincent}, {von Fellenberg}, {Ward-Duong}, {Widmann}, {Wieprecht},
  {Wiezorrek}, {Woillez}, \& {Gravity
  Collaboration}}]{nowakDirectConfirmationRadialvelocity2020}
{Nowak}, M., {Lacour}, S., {Lagrange}, A.~M., {et~al.} 2020, \aap, 642, L2

\bibitem[{{\"O}berg {et~al.}(2011){\"O}berg, {Murray-Clay}, \&
  Bergin}]{obergEffectsSnowlinesPlanetary2011}
{\"O}berg, K.~I., {Murray-Clay}, R., \& Bergin, E.~A. 2011, The Astrophysical
  Journal Letters, 743, L16

\bibitem[{Plummer(1918)}]{plummerIntroductoryTreatiseDynamical1918}
Plummer, H. C.~K. 1918, Cambridge

\bibitem[{{Rein} \& {Liu}(2012)}]{Rein2012}
{Rein}, H. \& {Liu}, S.~F. 2012, \aap, 537, A128

\bibitem[{{Rein} \& {Spiegel}(2015)}]{Rein2015}
{Rein}, H. \& {Spiegel}, D.~S. 2015, \mnras, 446, 1424

\bibitem[{{Snellen} \& {Brown}(2018)}]{2018NatAs...2..883S}
{Snellen}, I.~A.~G. \& {Brown}, A.~G.~A. 2018, Nature Astronomy, 2, 883

\bibitem[{{Straubmeier} {et~al.}(2014){Straubmeier}, {Yazici}, {Wiest}, {Wank},
  {Fischer}, {Eisenhauer}, {Perrin}, {Perraut}, {Brandner}, {Amorim},
  {Sch{\"o}ller}, \& {Eckart}}]{2014SPIE.9146E..29S}
{Straubmeier}, C., {Yazici}, S., {Wiest}, M., {et~al.} 2014, in Society of
  Photo-Optical Instrumentation Engineers (SPIE) Conference Series, Vol. 9146,
  Optical and Infrared Interferometry IV, ed. J.~K. {Rajagopal}, M.~J.
  {Creech-Eakman}, \& F.~{Malbet}, 914629

\bibitem[{{The Astropy Collaboration} {et~al.}(2018){The Astropy
  Collaboration}, {Price-Whelan}, {Sip{\H o}cz}, {G{\"u}nther}, {Lim},
  {Crawford}, {Conseil}, {Shupe}, {Craig}, {Dencheva}, {Ginsburg},
  {VanderPlas}, {Bradley}, {P{\'e}rez-Su{\'a}rez}, {de Val-Borro}, {Paper
  Contributors}, {Aldcroft}, {Cruz}, {Robitaille}, {Tollerud}, {Coordination
  Committee}, {Ardelean}, {Babej}, {Bach}, {Bachetti}, {Bakanov}, {Bamford},
  {Barentsen}, {Barmby}, {Baumbach}, {Berry}, {Biscani}, {Boquien}, {Bostroem},
  {Bouma}, {Brammer}, {Bray}, {Breytenbach}, {Buddelmeijer}, {Burke},
  {Calderone}, {Cano Rodr{\'{\i}}guez}, {Cara}, {Cardoso}, {Cheedella},
  {Copin}, {Corrales}, {Crichton}, {DAvella}, {Deil}, {Depagne}, {Dietrich},
  {Donath}, {Droettboom}, {Earl}, {Erben}, {Fabbro}, {Ferreira}, {Finethy},
  {Fox}, {Garrison}, {Gibbons}, {Goldstein}, {Gommers}, {Greco}, {Greenfield},
  {Groener}, {Grollier}, {Hagen}, {Hirst}, {Homeier}, {Horton}, {Hosseinzadeh},
  {Hu}, {Hunkeler}, {Ivezi{\'c}}, {Jain}, {Jenness}, {Kanarek}, {Kendrew},
  {Kern}, {Kerzendorf}, {Khvalko}, {King}, {Kirkby}, {Kulkarni}, {Kumar},
  {Lee}, {Lenz}, {Littlefair}, {Ma}, {Macleod}, {Mastropietro}, {McCully},
  {Montagnac}, {Morris}, {Mueller}, {Mumford}, {Muna}, {Murphy}, {Nelson},
  {Nguyen}, {Ninan}, {N{\"o}the}, {Ogaz}, {Oh}, {Parejko}, {Parley}, {Pascual},
  {Patil}, {Patil}, {Plunkett}, {Prochaska}, {Rastogi}, {Reddy Janga},
  {Sabater}, {Sakurikar}, {Seifert}, {Sherbert}, {Sherwood-Taylor}, {Shih},
  {Sick}, {Silbiger}, {Singanamalla}, {Singer}, {Sladen}, {Sooley},
  {Sornarajah}, {Streicher}, {Teuben}, {Thomas}, {Tremblay}, {Turner},
  {Terr{\'o}n}, {van Kerkwijk}, {de la Vega}, {Watkins}, {Weaver}, {Whitmore},
  {Woillez}, {Zabalza}, \& {Contributors}}]{2018AJ....156..123T}
{The Astropy Collaboration}, {Price-Whelan}, A.~M., {Sip{\H o}cz}, B.~M.,
  {et~al.} 2018, \aj, 156, 123

\bibitem[{{Th{\'e}bault} \& {Beust}(2001)}]{2001A&A...376..621T}
{Th{\'e}bault}, P. \& {Beust}, H. 2001, \aap, 376, 621

\bibitem[{Vandal {et~al.}(2020)Vandal, Rameau, \&
  Doyon}]{vandalDynamicalMassEstimates2020}
Vandal, T., Rameau, J., \& Doyon, R. 2020, AJ, 160, 243

\bibitem[{Vousden {et~al.}(2016)Vousden, Farr, \&
  Mandel}]{vousdenDynamicTemperatureSelection2016}
Vousden, W.~D., Farr, W.~M., \& Mandel, I. 2016, Monthly Notices of the Royal
  Astronomical Society, 455, 1919

\bibitem[{Wang {et~al.}(2018)Wang, Graham, Dawson, Fabrycky, De~Rosa, Pueyo,
  Konopacky, Macintosh, Marois, Chiang, Ammons, Arriaga, Bailey, Barman,
  Bulger, Chilcote, Cotten, Doyon, Duch{\^e}ne, Esposito, Fitzgerald, Follette,
  Gerard, Goodsell, Greenbaum, Hibon, Hung, Ingraham, Kalas, Larkin, Maire,
  Marchis, Marley, Metchev, {Millar-Blanchaer}, Nielsen, Oppenheimer, Palmer,
  Patience, Perrin, Poyneer, Rajan, Rameau, Rantakyr{\"o}, Ruffio, Savransky,
  Schneider, Sivaramakrishnan, Song, Soummer, Thomas, Wallace, {Ward-Duong},
  Wiktorowicz, \& Wolff}]{wangDynamicalConstraintsHR2018}
Wang, J.~J., Graham, J.~R., Dawson, R., {et~al.} 2018, The Astronomical
  Journal, 156, 192

\bibitem[{Wang {et~al.}(2021)Wang, Vigan, Lacour, Nowak, Stolker, De~Rosa,
  Ginzburg, Gao, Abuter, Amorim, {Asensio-Torres}, Baub{\"o}ck, Benisty,
  Berger, Beust, Beuzit, Blunt, Boccaletti, Bohn, Bonnefoy, Bonnet, Brandner,
  Cantalloube, Caselli, Charnay, Chauvin, Choquet, Christiaens, Cl{\'e}net,
  Coud{\'e} Du~Foresto, Cridland, {de Zeeuw}, Dembet, Dexter, Drescher, Duvert,
  Eckart, Eisenhauer, Facchini, Gao, Garcia, Garcia~Lopez, Gardner, Gendron,
  Genzel, Gillessen, Girard, Haubois, Hei{\ss}el, Henning, Hinkley, Hippler,
  Horrobin, Houll{\'e}, Hubert, {Jim{\'e}nez-Rosales}, Jocou, Kammerer,
  Keppler, Kervella, Meyer, Kreidberg, Lagrange, Lapeyr{\`e}re, Le~Bouquin,
  L{\'e}na, Lutz, Maire, M{\'e}nard, M{\'e}rand, Molli{\`e}re, Monnier,
  Mouillet, M{\"u}ller, Nasedkin, Ott, Otten, Paladini, Paumard, Perraut,
  Perrin, Pfuhl, Pueyo, Rameau, Rodet, {Rodr{\'i}guez-Coira}, Rousset,
  Scheithauer, Shangguan, Shimizu, Stadler, Straub, Straubmeier, Sturm,
  Tacconi, {van Dishoeck}, Vincent, {von Fellenberg}, {Ward-Duong}, Widmann,
  Wieprecht, Wiezorrek, Woillez, \& {Gravity
  Collaboration}}]{wangConstrainingNaturePDS2021}
Wang, J.~J., Vigan, A., Lacour, S., {et~al.} 2021, The Astronomical Journal,
  161, 148

\bibitem[{Wisdom \& Holman(1991)}]{wisdomSymplecticMapsNbody1991}
Wisdom, J. \& Holman, M. 1991, The Astronomical Journal, 102, 1528

\bibitem[{{Wolszczan} \& {Frail}(1992)}]{1992Natur.355..145W}
{Wolszczan}, A. \& {Frail}, D.~A. 1992, \nat, 355, 145

\end{thebibliography}

\end{document}